\documentclass[sigconf]{acmart}

\copyrightyear{2025}
\acmYear{2025}
\acmConference[FAccT '25]{The 2025 ACM Conference on Fairness, Accountability, and Transparency}{June 23--26, 2025}{Athens, Greece}
\acmBooktitle{The 2025 ACM Conference on Fairness, Accountability, and Transparency (FAccT '25), June 23--26, 2025, Athens, Greece}
\acmDOI{10.1145/3715275.3732022}
\acmISBN{979-8-4007-1482-5/2025/06}

\newcommand{\G}{\mathcal{G}}

\begin{document}

\title{Using Collective Dialogues and AI to Find Common Ground Between Israeli and Palestinian Peacebuilders}

\author{Andrew Konya}
    \orcid{0009-0003-6625-3457}
    \affiliation{%
      \institution{Remesh}
      \city{Cleveland}
      \state{OH}
      \country{USA}
    }
    \email{andrew@remesh.org}

\author{Luke Thorburn}
    \orcid{0000-0003-4120-5056}
    \affiliation{%
      \institution{King's College London}
      \city{London}
      \country{UK}
    }
    \email{luke.thorburn@kcl.ac.uk}

\author{Wasim Almasri}
    \orcid{0009-0002-6349-4514}
    \affiliation{%
      \institution{Alliance for Middle East Peace}
      \city{Ramallah}
      \country{Palestine}
    }
    \email{wasim@allmep.org}

\author{Oded Adomi Leshem}
    \orcid{0000-0002-8853-7819}
    \affiliation{%
      \institution{Hebrew University of Jerusalem}
      \city{Jerusalem}
      \country{Israel}
    }
    \email{oded.leshem@mail.huji.ac.il}

\author{Ariel D. Procaccia}
    \orcid{0000-0002-8774-5827}
    \affiliation{%
      \institution{Harvard University}
      \city{Cambridge}
      \state{MA}
      \country{USA}
    }
    \email{arielpro@seas.harvard.edu}

\author{Lisa Schirch}
    \orcid{0000-0003-4719-0473}
    \affiliation{%
      \institution{University of Notre Dame}
      \city{South Bend}
      \state{IN}
      \country{USA}
    }
    \email{lschirch@nd.edu}

\author{Michiel A. Bakker}
    \orcid{0000-0003-4474-7109}
    \affiliation{%
      \institution{Massachusetts Institute of Technology}
      \city{Cambridge}
      \state{MA}
      \country{USA}
    }
    \email{bakker@mit.edu}

\renewcommand{\shortauthors}{Konya et al.}

\begin{abstract}
    A growing body of work has shown that AI-assisted methods --- leveraging large language models, social choice methods, and collective dialogues --- can help navigate polarization and surface common ground in controlled lab settings. But what can these approaches contribute in real-world contexts? We present a case study applying these techniques to find common ground between Israeli and Palestinian peacebuilders in the period following October 7th, 2023. From April to July 2024 an iterative deliberative process combining LLMs, bridging-based ranking, and collective dialogues was conducted in partnership with the Alliance for Middle East Peace. Around 138 civil society peacebuilders participated including Israeli Jews, Palestinian citizens of Israel, and Palestinians from the West Bank and Gaza. The process resulted in a set of collective statements, including demands to world leaders, with at least 84\% agreement from participants on each side. In this paper, we document the process, results, challenges, and important open questions.
\end{abstract}




\maketitle


\section{Introduction}

    This paper documents an effort to find common ground between Israeli and Palestinian peacebuilders amidst deeply entrenched divisions that were amplified in the immediate aftermath of October 7th, 2023 and shaped by the broader context of protracted asymmetric conflict. From April to July 2024, an iterative deliberative process that combined collective dialogues \cite{ovadya2023generative}, bridging-based ranking \cite{ovadya2022bridging}, and large language models (LLMs) \cite{goldberg2024ai} was carried out in partnership with the Alliance for Middle East Peace (ALLMEP). More than 100 civil society peacebuilders participated in the process including Israeli Jews, Palestinian citizens of Israel, and Palestinians from the West Bank and Gaza, representing diverse lived experiences and interests. The process produced a set of collective statements (Table \ref{table:statements}) reflecting points of common ground --- including five statements to world leaders, and five to residents of the region --- with each statement supported by at least 84\% of peacebuilders on each side.

    In recent years, there have been several proposals using algorithmic methods to address polarization and find common ground. Bridging-based ranking, a class of recommendation algorithms that aims to surface content that builds trust and mutual understanding across divides \cite{ovadya2022bridging,ovadya2023}, was first implemented on Australian policy forum \textit{YourView} in 2012 \cite{van-gelder2012}, and later gained prominence when independently implemented on \textit{Polis} \cite{small2021polis,polisMathCommit8394f1f} and the Community Notes feature on X (then Twitter) \cite{wojcik2022bird}. Large-scale, online collective dialogues \cite{ovadya2023generative} that incorporate matrix completion methods to reduce the burden of preference elicitation have been used by the United Nations for peacebuilding since 2020 \cite{bilich2019faster,masood2022using,irwin2021using}. Most recently, there have been several proposals to use LLMs and social choice mechanisms to find common ground and make collective decisions on divisive issues \cite{bakker2022fine,small2023opportunities,fish2023generative,tessler2024ai,de2025supernotes}.
    
    Most study of these methods --- particularly bridging-based ranking and LLM-based approaches to articulating common ground --- has occurred in lab settings, offline, or for relatively low-stakes contexts. For example, the approach used in this case study builds on a specific integration of these methods that was initially developed to produce common ground policy as part of OpenAI's \textit{Democratic Inputs to AI} program \cite{konya2023dem}, and subsequently refined for use in a study on AI alignment \cite{konya2024chain}. In contrast, we present a case study using a combination of these methods to find common ground between cross-border stakeholders amid active armed violence. This required us to address a range of real-world challenges related to trust, legitimacy, language barriers, and a constantly-changing situation. We believe this work may be the first to examine the use of LLMs to bridge divides during a high-stakes real-world conflict.

    Given this context, it is important to note two principles that governed our approach:
    \begin{itemize}
    
        \item \textbf{Impartiality}:
        Drawing on best practices from peacebuilding \cite{un_peacekeeping_2008} and mediation \cite{gaffney2022impartiality}, we aimed for both our process and technology to be unbiased and fair: eliciting and representing participant perspectives without bias, and enabling equitable participation and influence over outcomes. 
    
        \item \textbf{Pragmatism}: Given a context which presents urgent challenges, we prioritized iterative approaches to identify viable solutions, rather than pursuing ideal outcomes that would hinder progress due to extensive development or refinement.
    
    \end{itemize}
    In this vein, we emphasize this paper presents a case study from a project aimed at supporting real-world peace processes, not academic research, which limited opportunities for controlled experimentation or formal evaulation. As such, our approach reflects a pragmatic reconciliation of established techniques with the constraints of a complex sociopolitical situation. Nonetheless, we believe the project can inform future academic work on the design and use of technology in conflict settings and are thus, with the permission of ALLMEP, publishing this case study.

    The paper is structured as follows. Section \ref{sec:background} reviews situational and technical context which motivated the project. Section \ref{sec:method} documents the method used to find common ground and explains  key design decisions. Section \ref{sec:results} reports the main insights and statements (representing common ground) that resulted from the process. Finally, in Section \ref{sec:discussion} (and related appendices) we discuss ethical considerations, the limitations of the process, and directions we intend to prioritize in future work.

    \begin{figure*}
        \centering
        \includegraphics[width=0.9\textwidth]{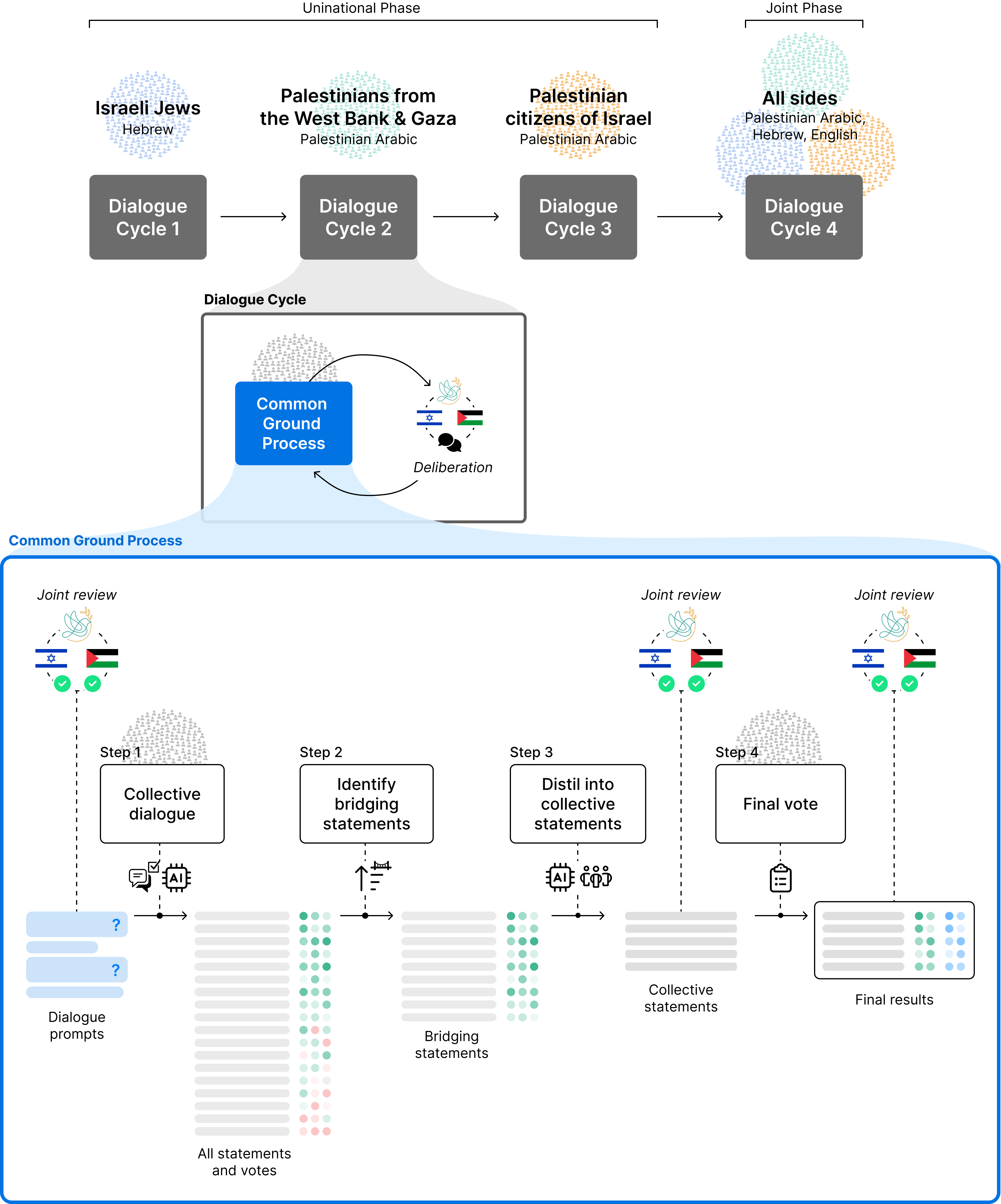}
        \caption{The effort consisted of four dialogue cycles with peacebuilders: three uninational dialogues cycles conducted respectively with Israeli Jews, Palestinians from the West Bank and Gaza, and Palestinian citizens of Israel, followed by a final joint dialogue involving all three groups. Each dialogue cycle involved a process to find common ground among participants followed by a deliberation on the results of that process. The common ground process started with a collective dialogue, then bridging-based ranking was used to identify common ground ``bridging statements'' that were distilled into articulate ``collective statements'' via LLM and reviewed by human experts before being shared back with participants for a final vote.
        The method is described in detail in Section \ref{sec:method}.}
        \label{fig:process}
    \end{figure*}

\section{Background}
\label{sec:background}

\subsection{Situational}
    
    In the context of geopolitical conflicts, ``track 1 diplomacy'' involving formal negotiations between governments is often supplemented or supplanted by ``backchannel'' dialogue involving unofficial non-governmental representatives from each side. Such backchannel dialogues may involve government officials participating in an unofficial capacity (``track 1.5 diplomacy'') or may not involve governments at all (``track 2 diplomacy'') \cite{montville1991track}. Often, these dialogues provide a more open environment to build trust and discuss tough issues. Furthermore, they can enable cross-border communication when formal diplomatic channels are closed. Civil society peacebuilders typically play a significant role in convening and contributing to these dialogues.

    In early 2024, following the events of October 7th, 2023, and within the broader context of protracted asymmetrical conflict, tensions between Israeli and Palestinian peacebuilders had understandably intensified. Shaped not only by the recent escalation in violence, but also by decades of structural disparity, open hostility, and deep-seated mistrust, the substrate to discuss tough issues had grown fragile\footnote{In this paper, we intentionally chose to use politically neutral language and not impose a characterization of the lived experiences of Palestinians and Israelis, including our own. This includes deliberately omitting certain terms, such as ``occupation,'' that have widely recognized legal and political significance under international law and are central to the lived experiences of many participants in this process. We believe it is consistent with the spirit of this work that any political positions conveyed by this otherwise-technical paper should not be ours, but those manifest in the points of common ground found among the peacebuilders who participated in this effort (Table~\ref{table:statements}).}. In this context, we worked with the Alliance for Middle East Peace (ALLMEP) --- a network of civil society peacebuilders in the region --- with the goal of strengthening that substrate. This initiative sought to create a foundation for meaningful dialogue grounded in shared humanity and collective  action. To this end, we sought to help peacebuilders find common ground. But the situation presented a few real-world challenges.
    
    \textbf{Language}: Peacebuilders' native languages span Hebrew, Palestinian Arabic, and English. Beyond communication, language carries deep cultural and historical significance, where it has been a tool of identity and a subtext of power imbalance. It was crucial for inclusivity and legitimacy that participants could engage fully in their native languages. Additionally, the sensitivity of language in this context extends to word choices that may reflect deeply rooted political and cultural meanings.
    
    \textbf{Trust}: Skepticism and mistrust, stemming from historical experiences of injustice, violence and prolonged conflict, posed significant barriers to engagement. This meant that before we could reasonably expect willingness to participate in joint dialogue, we would need to elevate trust in good faith partnership. It also meant that the process itself needed to hold up to scrutiny that the ``other side'' did not have a mechanism to influence the results to their advantage. 
        
    \textbf{Asymmetry}: We expected, by default, there would be an unequal number of representatives participating from each side, yet it would be anti-inclusionary to turn away participants from a side to achieve balance. It was crucial to design a process that accounted for structural asymmetries, ensuring that even with unequal representation, all ``sides'' held equitable influence on the outcomes.

\subsection{Technical}

\subsubsection{Collective dialogues}

    An online collective dialogue process is an iterative back-and-forth exchange between a moderator and participants. Participants join by clicking a link, and fill out demographic information as they enter the dialogue. During each turn of the dialogue, participants are sent either a read-only message, a standard survey question, or an open-ended prompt that kicks off a collective response process. During a collective response process~\cite{ovadya2023generative}, participants submit statements in response to the prompt, then evaluate statements submitted by others. The evaluation step serves two purposes: it exposes participants to others' views to help them better understand each other, and it elicits data for quantifying each statement's representativeness. On the collective dialogue platform used for this project (\textit{Remesh}), two types of evaluations are elicited; agreement votes, and pair choice votes. To maximize the number of dialogue turns viable within a fixed budget of participant time, and decouple votes per participant needed from the number of statements submitted, votes are sparsely sampled and a matrix completion method \cite{konya2022elicitation,bilich2019faster} is used to convert a sparse vote sampling into a complete participant vote matrix. The percent agreement each participant segment has with each statement is then computed by combining the complete vote matrix with the demographic data of the participants. Collective dialogues can be run both live and asynchronously (see Appendix \ref{appendix:collective}). They typically include on the order of ten collective response prompts, and take participants between 15-60 minutes to complete.

    The United Nations has run collective dialogues on Remesh --- often referred to as ``Digital Dialogues'' --- as part of peacebuilding efforts since 2020 \cite{futuringPeaceDigitalDialoguesYemen}. These collective dialogues are sometimes conducted by special political missions in support of track 1 diplomatic processes, as was done by Stephanie Williams during ceasefire negotiations in Libya \cite{UNLibya,irwin2021using}. In other cases, they serve less formal track 1.5 or track 2 activities associated with local field presences. Overall, the UN has now run more than 60 collective dialogues across the world including in Yemen \cite{futuringPeaceDigitalDialoguesYemen}, Iraq \cite{UNIraq}, Lebanon \cite{UNLebanon}, Haiti \cite{UNHati}, Bolivia \cite{UNBolivia,UNBolivia2}, and Bahrain \cite{UNBahrain}.

\subsubsection{Bridging-based ranking}
\label{sec:bridging-based-ranking}

    The term bridging-based ranking refers to a class of ranking or recommendation algorithms that aim to ``increase mutual understanding and trust across divides, creating space for productive conflict, deliberation, or cooperation''`\cite{ovadya2023}.  Most commonly, this qualitative goal is operationalized as ``diverse approval,'' meaning that items are ranked by the degree to which they are approved of or valued by clusters of participants who would normally be expected to disagree with each other (in other words, ranking by diverse approval surfaces a form of ``common ground''). Implementations of this core intuition differ, for example with respect to whether clusters are stipulated or learned from the data \cite{small2021polis}, and how diversity is quantified. Below, we review two existing approaches which we used in this project.
    
    Let $P$ denote a set of people, $\G$ denote a set of groups (that is, a set of subsets of $P$), and $S$ denote a set of items to be ranked, which in our context are natural language statements. Perhaps the simplest approach is \textbf{max-min agreement}, where items are ranked by the minimum rate of agreement they receive across each group. Formally, for a group $G \in\G$ and statement $s\in S$, let $a(s,G)$ denote the proportion of people in group $G$ that agree with statement $s$. Under the max-min agreement approach, statements in $s\in\mathcal{S}$ are ranked according to the metric
    $$a_\textsf{min}(s,\G) = \min \{a(s,G)\mid G\in\G\}.$$
    This metric has been used previously in collective dialogue processes run on Remesh \cite{konya2023dem,konya2024chain}.
        
    Another prominent approach to bridging is the algorithm used by the \textbf{Community Notes} feature on the social media platform X. It elicits crowd-sourced notes that add context to posts, but only shows them publicly if (to simplify slightly) they are rated as helpful by people on both sides of a learned political spectrum. 
    To achieve this, the algorithm fits a latent factor model to the observed agreement votes, which provides for each note a factor corresponding to its slant along the most salient dimension of disagreement \cite{wojcik2022bird}. Formally, let $v(s,p)\in\{1,-1\}$ denote whether person $p$ said they respectively agreed or disagreed with statement $s$. The Community Notes algorithm then fits the following model
    $$\hat{v}(s,p) \approx \mu + i_p + i_s + \mathbf{f}_p \cdot \mathbf{f}_s.$$
    The global intercept $\mu$ captures the overall likelihood of any voter to agree\footnote{In the Community Notes implementation used for posts on X, the primary vote reflected in $\hat{v}(s,p)$ is not ``agree'', but ``helpful''. While the math is the same, here we describe the algorithm in terms of ``agree'' votes in order to maintain consistency across descriptions of the methods used in this paper.} with \emph{any} statement, the user intercept $i_p$ captures voter $p$'s overall likelihood to agree with \emph{any} statement, and the note intercept $i_s$ captures the voting population's overall likelihood to agree with statement $s$. The last term $\mathbf{f}_p \cdot \mathbf{f}_s$ represents the product of the voter and statement factors,\footnote{These factors can be scalars that capture only the primary axis of voter variance, as in the original Birdwatch paper \cite{wojcik2022bird}, or vectors that capture multi-dimensional factors of vote variance.}, respectively. Model parameters are fit by minimizing a regularized least-squared loss via gradient descent over the dataset of observed votes. Once learned, the note and user factors map statements and voters along learned axes spanning ``sides'' of division, while the note intercept $i_s$ gives a side-independent measure of agreement. Because the note intercept $i_s$ is side-independent, we can use it to find statements that bridge across people on different sides.
    
\subsubsection{Language models}

    Large language models (LLMs) are increasingly being recognized for their potential to address societal challenges by enhancing collaboration and fostering collective intelligence. These models offer new ways to process and synthesize diverse viewpoints, enabling the articulation of shared perspectives even in deeply divided groups \cite{argyle2023leveraging,fish2023generative,tessler2024ai}. For example, in collective dialogue systems, LLMs can distill participant responses into coherent statements that reflect areas of agreement. Beyond synthesizing viewpoints and identifying areas of agreement, LLMs can play a critical role in overcoming language barriers. They offer real-time translation, text-to-speech and summarization capabilities, ensuring participants can engage fully in their native languages. This functionality helps make deliberative processes more inclusive, equitable, and accessible \cite{burton2024large}. However, careful implementation is essential to avoid risks such as bias and misrepresentation.

\section{Method}
\label{sec:method}

\subsection{Common ground process}

    The core of our approach is a process (Figure \ref{fig:process}) that integrates collective dialogues, bridging-based ranking, and LLMs\footnote{For all LLM tasks in this project, we used models from OpenAI's GPT-4 series. Specifically, during the uninational phase we used \texttt{gpt-4-turbo-2024-04-09}, and during the joint phase we used \texttt{gpt-4-0125-preview}.} to help identify, articulate, and validate points of common ground among a group of participants. The process begins with a collective dialogue (Step 1) in which participants anonymously respond to prompts and evaluate the responses of others. The responses that are most likely to represent common ground are identified using bridging-based ranking (Step 2) and distilled into a set of well-articulated collective statements using an LLM pipeline (Step 3). The collective statements are then shared back with participants for a final vote (Step 4). The primary outputs of each run of the process are a set of collective statements along with the results of the final vote. We describe each step in more detail below.

\subsubsection{Step 1: Collective dialogue}

    The first step towards a collective dialogue is drafting the planned prompts participants will respond to and deciding what demographics to gather. Prompt drafting is informed by the goals of the dialogue: specific issues to find common ground on, areas to create more mutual understanding around, and so on. Draft prompts are reviewed and iterated until deemed acceptable by a stakeholder group with representation from all sides. The choice of demographics are determined by (a) demographic dimensions likely to manifest relevant divisions, (b)  demographics needed to assess representativeness, and (c) constraints imposed by the need for statistical anonymity \cite{bravohermsdorff2022statistical}.

    In this work, the collective dialogues were first run live, then reopened for asynchronous participation as needed. The date and time of the live dialogue was typically set a week or two in advance and communicated with participants so they could plan accordingly. The live event began with a Zoom video call where the dialogue was contextualized, then a link was shared with participants to join the collective dialogue on Remesh. After the live dialogue was over, it was re-opened for asynchronous participation for a few days. 

    Each collective dialogue generates a a set of natural language statements $S$ responding to each prompt, a sparse sampling of votes on each statement $s \in S$, participant demographics, and data for the non-collective-response questions asked during the dialogue. Vote inference \cite{konya2022elicitation} is then used to complete the agreement-vote matrix with elements $v(s,p) \in \{1,0\} \equiv \{agree,disagree\}$ that indicate the (stated or inferred) agreement of participant $p$ with statement $s$.

\subsubsection{Step 2: Identifying bridging statements}

    The goal of this step is to identify a set of ``bridging'' statements $B\subseteq S$ that are likely to represent points of common ground. To do this, we computed two bridging metrics for each statement $s\in S$: (1) max-min agreement $a_\textsf{min}(s,\G)$ across defined demographic groups $\G$, and (2) the learned parameters $(f_s,i_s)$ from the Community Notes algorithm (see Section \ref{sec:bridging-based-ranking}). As a first set of bridging statements, we selected statements $s\in S$ where $a_\textsf{min}(s,\G)$ was greater than a given threshold. This gives the set of statements that have the highest lowest agreement across the specified set $\G$ of demographic groups. However, when $G$ contains many demographic groups --- defined by age, gender, religiosity, etc. --- this approach may exclude a statement that may otherwise bridge a critical axis of division just because it is predicted (perhaps incorrectly) to have low agreement among one demographic sub-group. To mitigate this risk, we created a second set of bridging statements using the Community Notes approach by selecting those where the Euclidean distance $d((f_s,i_s),(0,2))$ was below a given threshold. This approach relies primarily on the note intercept $i_s$ as a bridging metric, but also penalizes statements for having non-zero slant, as indicated by the note factor $f_s$. The complete set of bridging statements $B$ is taken to be the union of the sets of statements emerging from both approaches.
    
\subsubsection{Step 3: Distillation into collective statements}

    The set of bridging statements identified in the previous step typically includes multiple restatements of the same idea, compound statements covering multiple ideas, and sub-optimal articulations of ideas. The goal of Step 3 is to create well-articulated versions of each of unique idea that appears in the bridging statements in a way that preserves participants' choices of specific words and phrases. We use an LLM to do this in two steps. First, the LLM is prompted to extract the unique ideas from the set of bridging statements. Next, the LLM is prompted to generate well-articulated versions of each unique idea in a way that preserves the words and phrases found in the original statements. Included in this prompt are the set of unique ideas, the text of the bridging statements the unique ideas were derived from, and a set of exemplar statements that demonstrate the style in which collective statements should be written. Finally, the collective statements proposed by the LLM are reviewed by local language experts for linguistic consistency with the original bridging statements, and iterated as needed until a final set of collective statements $C$ is deemed fair by a stakeholder group with representation from all sides. The LLM prompts used for this step are included in Appendix \ref{appendix:prompts}.    

\subsubsection{Step 4: Final vote}

    We expect the collective statements that reach this point to generally represent common ground. However, the use of both AI and human experts in the prior steps introduce potential sources of error and bias. These risks can rightfully delegitimize the claim that the collective statements actually represent common ground. The final vote is thus a critical step that serves to strengthen legitimacy by measuring support for the collective statements in a simple and transparent way. During the final vote, all participants vote on all collective statements. There is no AI inference or synthesis, and results are computed via deterministic and reproducible methods.

    Participants cast two types of votes on each final collective statement. First, a simple vote of agreement with each statement is cast by each participant on a 5-point Likert scale. This enables a direct measurement of how well each statement collectively represents a point of agreement on all sides. In the second type of vote, participants are asked to rank the set of statements in order of agreement. Even if all collective statements represent clear points of common ground, they may not be equally important. The relative votes provide an additional signal to understand how the set of common ground collective statements should be prioritized.

\subsubsection{Joint reviews}

    Even though the final vote serves as a hedge against biases that could cause collective statements to not represent common ground, unfair bias at any point in the process can still undermine legitimacy. To mitigate this risk, at key junctures in the process where bias might be introduced, joint reviews were conducted by representatives of each side before moving on to the next step (see Figure \ref{fig:process}). Specifically, joint reviews were conducted on: the content and wording of all messages, questions, and collective response prompts that would be seen by participants in the collective dialogue, the collective statements that would go on to the final vote, and the results computed from the final vote data.

\subsection{Dialogue cycles}

\subsubsection{Transitional deliberation}
    We made our approach iterative by adding a transitional deliberation that marks the end of one dialogue cycle and the beginning of the next; serving to translate the outputs of one cycle to the inputs of the next cycle. These deliberations involved a small group of Israeli and Palestinian process facilitators and stakeholders. Each deliberation started with contextualization and sensemaking to establish mutual understanding related to the results, then transitioned to prompt design and participant communications for the next cycle.

\subsubsection{Strategic sequencing}
    Our goal was to run a joint common ground process involving cross-border dialogue between peacebuilders from all sides, but before doing so we sought to build trust in the process and among those involved. To this end, we began with a uninational phase, where each dialogue cycle involved participants from only one side, in order to first find points of common ground within each group that could be communicated to help elevate trust and willingness to engage in the subsequent cycles.

    What constitutes a ``side''? It may seem obvious that ``Israelis'' and ``Palestinians'' would be the two sides. However, this binary approach would have risked marginalizing Palestinian citizens of Israel, may not fully identify with either Israelis or Palestinians. To accommodate this, we chose to treat Palestinian citizens of Israel as a third independent group, along side Israeli Jews and Palestinians from the West Bank and Gaza. In the remainder of this paper, we use $\G$ to denote this set of groups, namely $\G =\{$Israeli Jews, Palestinian citizens of Israel, Palestinians from the West Bank and Gaza$\}$.

    During the uninational phase, the group that went first needed to do so without the benefit of any trust-building feedback from earlier cycles. Given the existing power asymmetry, it was decided that Israeli Jews would go first. As the second major group, Palestinians from the West Bank \& Gaza went second, followed by Palestinian citizens of Israel. Finally, after these three uninational dialogues, we conducted a joint dialogue with participants from all groups to attempt to identify cross-border common ground (Figure \ref{fig:process}).

\subsection{Participants and Experts}

    Dialogue participants were recruited via invitations circulated to ALLMEP member organizations (see, e.g. Appendix \ref{appendix:invite}). ALLMEP leadership also personally contacted leaders of different organizations to invite participation. 
    Because participants were anonymous %
    and there were multiple stages within each dialogue cycle with no way of tracking identities across stages, it's not possible to know the exact number of unique participants. We estimate a lower bounds on the number of unique, meaningfully engaged participants in each of the four dialogue cycles to be 70, 28, 13, and 138, respectively (see Appendix \ref{appendix:counts} for a detailed breakdown). 
    Additionally, the process included ``human experts'' in-the-loop, mentioned throughout this Methods section. This small group of experts was hand selected by process facilitators. They consisted of ALLMEP staff and two academics with relevant peacebuilding expertise. 

\subsection{Languages}

    It was critical for legitimacy and inclusivity that participants could fully engage in their primary languages. The uninational cycles would be mono-lingual (Hebrew or Palestinian Arabic), and the joint cycle would need to be multilingual. The main challenge for the mono-lingual cycles was the LLM pipeline used to generate collective statements. Early versions of the pipeline used in prior work \cite{konya2023dem,konya2024chain} had proven effective when the prompts, examples, bridging statements, and generated outputs were all in English, but it was unclear how to approach Hebrew and Palestinian Arabic. Should prompts or examples still be in English? And who decides? To pragmatically address these questions we set up two micro-experiments. We compiled a set of plausible bridging statements in each language and asked native speakers with prompt engineering experience to translate the English prompts and examples. Different variations of the pipeline were created by swapping out different permutations of the English prompts and examples with translated versions. Collective statements were generated using each variation. Then, blind to how they were generated, native speakers were asked which set of collective statements best reflected the content, culture, and language of the original statements they were based on. In all cases, reviewers preferred the statements generated by a pipeline with prompts and examples in their native language, so these variants were used for the uninational phase. 

    For the joint cycle, a trilingual approach encompassing Hebrew, Palestinian Arabic, and English was implemented to maximize accessibility and inclusivity.  This required the provision of all textual materials --- including collective response prompts, participant-generated statements, synthesized collective statements, and result presentations --- in all three languages. All collective dialogue prompts were initially drafted in English, then translated to the other languages using the Google Cloud Translation API (hereafter, ``Google Translate'') and refined by native speakers. During the live dialogue, Google Translate was used to create trilingual versions of each statement in real-time. Acknowledging the inherent limitations of machine translation and the potential for semantic nuances to be lost or distorted, we adopted a transparent approach, explicitly informing participants about the use of Google Translate. Participants were also asked to use clear and concise language to mitigate the risk of translation errors. We conducted a set of micro-experiments on the multilingual LLM pipeline similar to those run for the monolingual pipeline. The final LLM output was a set of trilingual collective statements that would be reviewed and refined by native speakers for consistency across translations  and  with the original bridging statements. Final results were compiled in a spreadsheet and a trilingual lookup table for all elements in main view results, including labels and metrics, was initialized via Google Translate and then jointly reviewed and refined by native speakers. 

\subsection{Equal power metrics}

    For the joint cycle, it was critical that result metrics computed from final vote data (ranking and agreement votes) give each side equal influence even with unequal shares of participants. To that end, we computed three equal-power metrics --- metrics that are invariant to the portion of participants from each side --- for each collective statement: max-min agreement, Dowdall score, and IRV rank.
    
\subsubsection{Max-min agreement}
    
    Max-min agreement, the same metric used for bridging-based ranking in Step 2, was computed again using the final agreement votes. Specifically, for each collective statement $s$, max-min agreement was computed as $\min\{ \alpha(s,G)\mid G \in \G\}$ where $\alpha(s,G)$ is the fraction of participants in group $G$ that voted ``agree'' or ``strongly agree'' on statement $s$. This metric inherently provides equal power across groups and offers a straightforward interpretation: it indicates the minimum level of agreement across all defined groups for statement $s$.

    \begin{table*}
        \caption{Top three \emph{outgroup} red lines along with potentially related \emph{ingroup} red lines from the other side. Note that the nature of relatedness between the red lines shown is non-specific, and does not imply direct alignment nor suggest equivalency in content or scope. In the column headers of this table, the adjectives  ``Israeli''  and ``Palestinian'' refers to participants of the uninational dialogue cycles with Israeli Jews and  Palestinians from the West Bank and Gaza, respectively.}
        \label{table:red-lines}
        \vspace{-3mm}
        \includegraphics[width=0.99\textwidth]{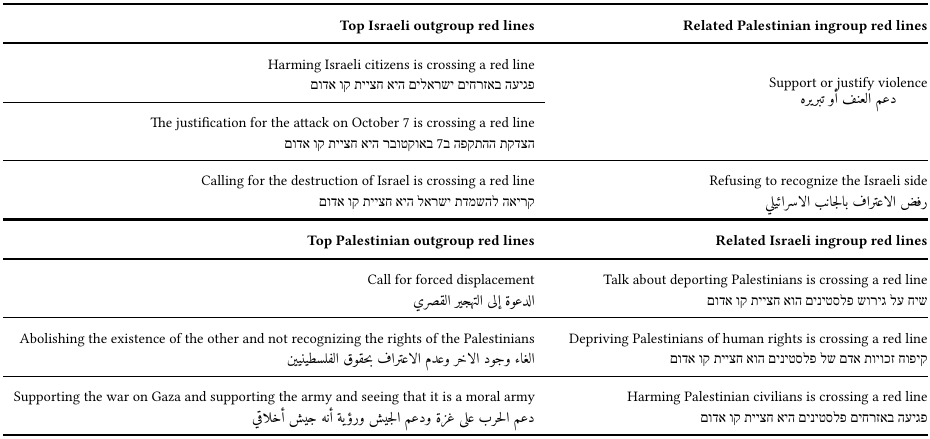}
    \end{table*}

    \begin{table*}
        \caption{Coarse-grained shared values and visions emerging from collective statements produced during the uninational phase. Note that the succinct shared value or vision expressed in each row does not fully entail all ideas contained the associated statements and only captures their abstract common thread. Similarly, presenting statements as associated with the same shared value or vision does not imply equivalency between them.}
        \label{table:values-and-visions}
        \vspace{-3mm}
        \includegraphics[width=0.99\textwidth]{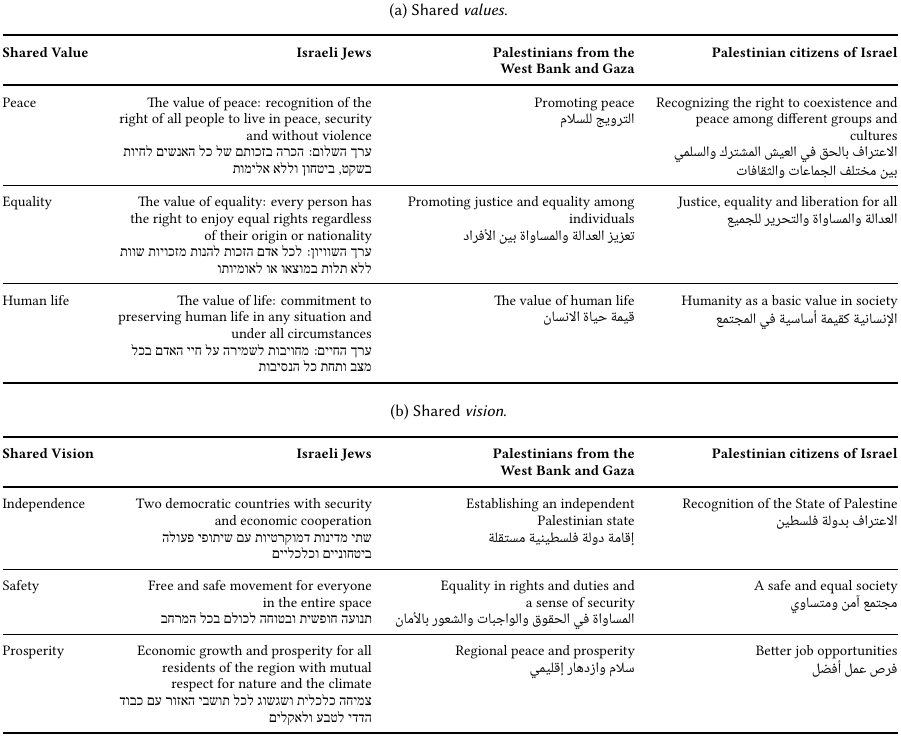}
    \end{table*}

    \begin{table*}
        \caption{Collective statements generated during the joint phase along with final vote results.
        }
        \label{table:statements}
        \vspace{-3mm}
        \includegraphics[width=.99\textwidth]{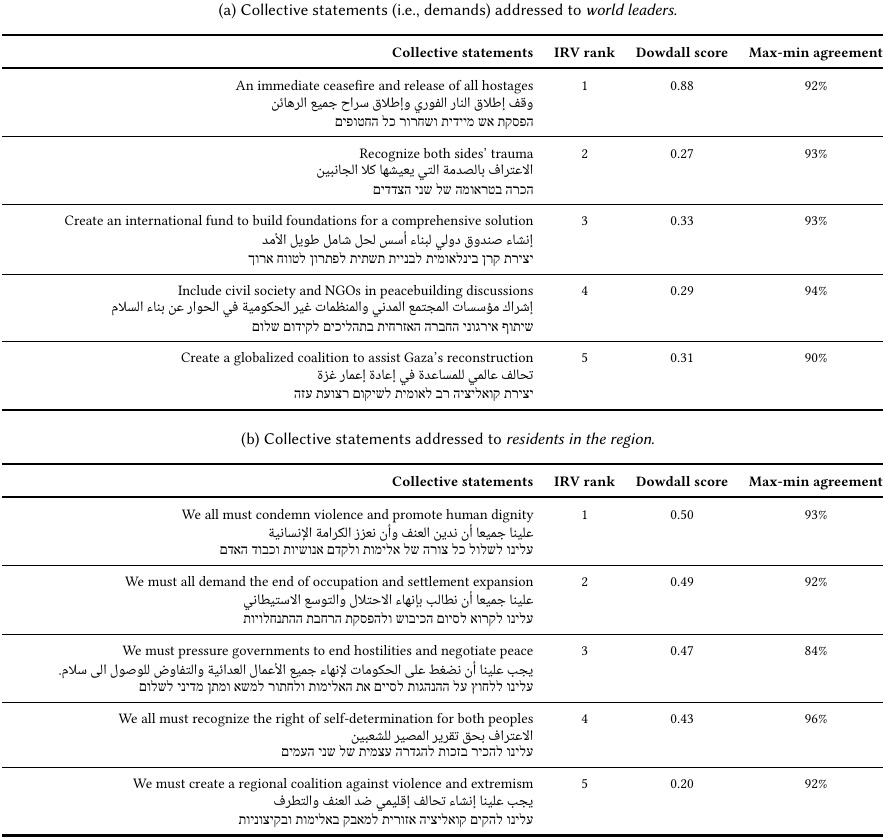}
    \end{table*}
 
\subsubsection{Dowdall score}

    Similar to a Borda count \cite{brandt2016handbook}, a Dowdall score is a cardinal metric computed from rank votes using a positional scoring rule that quantifies the strength of relative preference among a set of alternatives for a voting population \cite{fraenkel2014borda}. Let the fraction of participants in population $P$ who rank statement $s$ in position $k$ be denoted $\textsf{Position}(s,k;P)$. A Dowdall score is computed as
    $$\textsf{Dowdall}(s;P) = \sum_k \textsf{Points}(k) \times \textsf{Position}(s,k;P),$$
    where $\textsf{Points}(k)=1/k$ corresponds to the ``points'' assigned to each position. We adapt this to be an equal power metric by taking the average of Dowdall scores computed independently for each side:
    $$\textsf{Dowdall}_\textsf{EP}(s) = \frac{1}{|\G|} \sum_{G\in\G} \textsf{Dowdall}(s;G).$$
    This averaging approach ensures that each group's collective preferences contribute equally to the final equal-power Dowdall score, regardless of group size.
    
\subsubsection{IRV rank}

    An instant runoff voting (IRV) rank is an ordinal metric derived from rank votes by iteratively eliminating the option with the fewest first-place votes among remaining options \cite{brandt2016handbook}. Let $S_n$ be the set of statements remaining in round $n$, and $\textsf{First}(s,n;P)$ give the fraction of participants (in some population $P$) who rank statement $s$ first among $S_n$. The statements remaining in round $n+1$ are $S_{n+1} = S_n \setminus \{\arg\min_{s \in S_n} \textsf{First}(s,n;P) \}$.\footnote{This formulation assumes no ties for last place, which was true with our data.} Denoting $n_s$ to be the round where statement $s$ was eliminated, and $|C|$ the total number of statements being ranked, the IRV rank is $\textsf{IRV}(s) = |C| - n_s + 1$. To make this an equal power metric, the ``first place fraction'' $\textsf{First}(s,n)$ used to eliminate statements in each round is computed as an average of the first place fractions within each group. That is,
    $$\textsf{First}_\textsf{EP}(s,n) = \frac{1}{|\G|} \sum_{G\in\G} \textsf{First}(s,n;G).$$

\section{Results}
\label{sec:results}

    When interpreting the results presented here, it's important to consider a few contextual factors. First, given the sensitive nature of the issues addressed in these dialogues and their direct relevance to ongoing peacebuilding efforts, the results shared are intentionally selective. Specifically, we present: (1) findings from the uninational phase that demonstrably facilitated the subsequent joint phase, and (2) collective statements of common ground from the joint phase that were explicitly intended for external communication. Furthermore, the quantitative results we share are limited to metrics that give equal power representation across primary participant groups. Second, these findings should not be understood as the views of the broader Israeli and Palestinian populations. The participant pool, with an approximate sample size of no more than $N=138$ for any dialogue cycle, is not only statistically insufficient for generalization, but was comprised exclusively of civil society peacebuilders. Third, while some of the identified areas of common ground may appear self-evident in retrospect, at that time this project us undertaken, a prevailing sentiment held that finding common ground was unlikely, and the prospect of a constructive joint dialogue was viewed with considerable skepticism.

\subsection{Uninational Phase}

    A key aim of the uninational phase was building trust and creating the context under which a productive cross-border engagement in the joint phase could take place. To this end, each cycle was designed to identify \emph{red lines}, \emph{values}, and \emph{visions} for the future. 

\subsubsection{Red lines}

    In our work, the term ``red lines'' refers to actions (including speech acts) that those participating in the process believe are unacceptable, and if undertaken may prevent further dialogue. We asked about two kinds of red lines: \textit{ingroup} (the lines that each side themselves would not cross) and \textit{outgroup} (the lines that each side would not accept the ``other side'' crossing). These red lines served a few purposes. Ingroup red lines would provide a trust building signal if they articulated positions that demonstrated respect for the concerns of the other side. The combination of both types of red line across groups would help clarify the boundary conditions under which a productive joint dialogue could take place. In this context, one notable result of this phase was the mutual alignment between the ingroup and outgroup red lines of Israeli Jews and Palestinians from the West Bank \& Gaza. For each of the top three outgroup red lines on each side, the other side had an arguably related ingroup red line that demonstrated some degree of relevant respect and aligned ethics (Table \ref{table:red-lines}). For example, ``call for forced displacement'' was a top outgroup red line from Palestinians from the West Bank \& Gaza, while ``talk about deporting Palestinians'' was a top ingroup red line from Israeli Jews. 

    Another notable result came during the transitional deliberation at the end of first dialogue cycle with Israeli Jews. While reviewing a top outgroup red line – ``the justification for the attack on October 7th'' – strong concern emerged that prohibiting explanation about why October 7th happened would be an unfair constraint on joint dialogue. In the exchange that followed, a mutual understanding emerged:  ``justification'' specifically involved a claim of moral rightness, while an ``explanation'' was a neutral analysis of causation. So claiming the attack was morally right was the red line, but explaining what caused it to happen was not. This distinction not only applied to many collective statements throughout the process, but helped introduce language like \emph{``I won’t justify [X], but an explanation is…''} as a rhetorical device for discussing tough issues. 

\subsubsection{Values and visions}
    
    Values and visions were intended to help each side articulate the moral values they held and the future outcomes they sought to achieve. In the spirit of a ``GRIT'' (Graduated and Reciprocated Initiatives in Tension reduction) approach \cite{lindskold1978trust}, our hypothesis was that commonalities would emerge across all sides that could serve as early points of common ground, prior to the joint dialogue process actually taking place. After all three uninational cycles had concluded, the collective statements entailing values and vision were compared and a set of coarse-grained shared values and visions (Table \ref{table:values-and-visions}) were identified: \textit{peace}, \textit{equality}, \textit{human life}, \textit{independence}, \textit{safety}, and \textit{prosperity}.

    Overall, around 100 peacebuilders participated in the uninational phase and the intra-group points of common ground that were identified were communicated back with peacebuilders on all sides. These results helped create the context for the joint phase to take place. The mutually compatible red lines helped restore some trust, and clarified the boundary conditions for a productive joint dialogue. The values and visions served as a reminder of the shared foundation that already existed and could be built on.

\subsection{Joint phase}
    
    The joint phase covered a range of topics. Here we discuss the motivation and results for two of those: demands to world leaders, and messages to residents in the region. These were chosen by working backward from what ALLMEP and its member organizations were able to action. Internationally, ALLMEP acts as an interlocutor between Israeli/Palestinian civil society and governments, and an amplifier in the media. So, if the process could generate a set of joint statements directed at world leaders, then ALLMEP would be in a position to communicate these through their ongoing advocacy work. And since ALLMEP's membership is composed of civil society peacebuilding organizations in the region, a set of joint messages to residents in the region could be used as part of their members public communications strategy. 

    Around 138 people participated in the joint phase. Five collective statements comprising joint demands to world leaders were produced (Table \ref{table:statements}a). Each statement had at least 90\% agreement among participants from each side (i.e., $a_\textsf{min}(s,\G)>90\%$), and one demand was universally ranked first: ``An immediate ceasefire and release of all hostages.'' These demands were then composed into a ``joint letter'' for inclusion in briefs and advocacy efforts (Appendix \ref{appendix:letter}). Additionally, five collective statements comprising joint messages to residents in the region were produced (Table \ref{table:statements}b). Each of these statements had at least $84\%$ agreement among participants from each side, with most statements being preferred about equally.

\section{Discussion}
\label{sec:discussion}

    In this section, we discuss how this work can be evaluated, and considerations and challenges encountered when using AI in this context. Further discussion of ethical issues, limitations, and future directions are included in Appendix \ref{appendix:discussion}.

\paragraph{Evaluation}

    As this project was conducted in support of an active peace process, it lacked the structure of well-controlled trials that support rigorous comparative evaluations. Further, our process did not directly replace an existing one to which it could be directly compared. To the extent that it was replacing anything, it could be partly viewed as replacing long deliberations where professional mediators employ standard, labour-intensive approaches to finding common ground, e.g., \cite{moore2014,dillard2013,schirch2015}. In contrast, ALLMEP noted that ``\emph{Without this process, our conversations would be smaller and fragmented. This method brought voices together across political and geographic divides in a way that traditional dialogue simply couldn't.}''

    The 10 public-oriented statements produced by this process had between 84-96\% bridging support, and nine had support over $90\%$. While no perfect analogue exists to say how ``good'' these levels of support are, we note that citizens assemblies --- often cited for their ability to find common ground on divisive issues --- generally produce recommendations that have 60-95\% overall support under less tense circumstances \cite{BCAssembly2004,Farrell2015MarriageEquality,BuergerRat2020, CarbonBrief2020, GlobalAssembly2022, Involve2018}.

    Participants in the process reported that the experience fostered "\emph{a lot of self-reflection}," that the anonymity enabled them to "\emph{share openly, without fear of being judged}" and that "\emph{seeing each other’s anonymous thoughts in this way is valuable in and of itself}." However, while most participants reported a positive experience, some thought that in the live collective dialogues "\emph{Everything is so fast. It’s hard to think.}" and that they "\emph{need to be able to speak to each other, see each other’s faces}'' (see Appendix \ref{appendix:feedback} for more of this feedback).

\paragraph{Use of AI}

    LLMs played a few critical roles in enabling this process. First, they rapidly distilled numerous raw statements into manageable, nuanced collective statements, replacing complex multi-stakeholder deliberations (which can take months) with an LLM-accelerated step of a few hours. This compressed each dialogue cycle to weeks, making a recursive process viable. Second, LLMs and machine translation made the multilingual cycle practical by reducing complex translation burdens, enabling multilingual collective statement synthesis, and facilitating live multi-lingual dialogues through low-latency translation. Third, LLM-enabled vote prediction \cite{konya2022elicitation} improved data efficiency for the bridging-based ranking and decreasing how much participants had to vote per prompt and thus allowing more prompts within time-constrained dialogues.

    A key challenge of this work was figuring out how to reap these benefits of using AI without sacrificing trust and legitimacy due to real/perceived issues with AI. Vote prediction is imperfect, LLMs can hallucinate, and machine translation is sometimes just wrong. Such imperfections hurt trust and legitimacy in a system that uses AI. However, rather than ensuring all AI used was unbiased, and error-free, our approach was designed under the assumption that all AI used would be imperfect. The process was formulated to accommodate AI imperfection by incorporating expert humans in the loop, being transparent with participants (Appendix \ref{appendix:transparency}), and using AI-free final votes to validate all points of common ground. In particular, all outputs of AI/LLMs were verified with humans-in-the-loop, with the exception of real-time machine translation during the dialogues for which we had the experts review sample translations in advance to choose the most accurate translation method available. AI was only used for tasks where it was faster or cheaper to have the AI do it and then have humans check the outputs, than to have the humans do it themselves.

\section{Conclusion}

    In this case study, we documented a recursive deliberative process in which we used online collective dialogues, bridging-based ranking, and LLMs to surface common ground between Israeli and Palestinian peacebuilders. While prior work has shown the effectiveness of integrating similar techniques in low-stakes settings, this work is an ``existence proof'' that (i) these technologies can be used within high-stakes conflict settings while maintaining trust; (ii) that a scalable method can successfully bridge divides among distrustful groups amid active conflict; and (iii) that such processes can be made equitable despite significant structural asymmetries between participants.


\clearpage

\begin{acks}

    We first thank the team at ALLMEP that worked extremely hard to make this effort happen: Doubi Schwartz, Wasim Almasri, Nivine Sandouka, Brian Reeves, Nick Acosta, Kari Reid, Dr. Natali Levin-Schwartz, John Lyndon, and Avi Meyerstein; Avi and John for their leadership and commitment to advancing this initiative during challenging times. We thank Hope Schroeder, Manuel Wuthrich, Karmel Salah, Jay Baxter, Colin Megill, and Colin Irwin for helpful conversations and input. Finally, we thank Jake Weiner, Ivy Cheung, Amanda Yonce, Margo Flewelling, Jennifer Gong, and Emmet Hennessy for rapidly developing the multi-lingual capability on Remesh that enabled timely execution of the joint phase.

    Luke Thorburn was supported in part by UK Research and Innovation [grant number EP/S023356/1], in the UKRI Centre for Doctoral Training in Safe and Trusted Artificial Intelligence (safeandtrustedai.org), King's College London.
    
\end{acks}

\bibliographystyle{ACM-Reference-Format}
\bibliography{references}


\begin{thebibliography}{45}


\ifx \showCODEN    \undefined \def \showCODEN     #1{\unskip}     \fi
\ifx \showDOI      \undefined \def \showDOI       #1{#1}\fi
\ifx \showISBNx    \undefined \def \showISBNx     #1{\unskip}     \fi
\ifx \showISBNxiii \undefined \def \showISBNxiii  #1{\unskip}     \fi
\ifx \showISSN     \undefined \def \showISSN      #1{\unskip}     \fi
\ifx \showLCCN     \undefined \def \showLCCN      #1{\unskip}     \fi
\ifx \shownote     \undefined \def \shownote      #1{#1}          \fi
\ifx \showarticletitle \undefined \def \showarticletitle #1{#1}   \fi
\ifx \showURL      \undefined \def \showURL       {\relax}        \fi
\providecommand\bibfield[2]{#2}
\providecommand\bibinfo[2]{#2}
\providecommand\natexlab[1]{#1}
\providecommand\showeprint[2][]{arXiv:#2}

\bibitem[Alavi et~al\mbox{.}(2022)]%
        {irwin2021using}
\bibfield{author}{\bibinfo{person}{Daanish~Masood Alavi}, \bibinfo{person}{Martin Wählisch}, \bibinfo{person}{Colin Irwin}, {and} \bibinfo{person}{Andrew Konya}.} \bibinfo{year}{2022}\natexlab{}.
\newblock \showarticletitle{{Using Artificial Intelligence for Peacebuilding}}.
\newblock \bibinfo{journal}{\emph{Journal of Peacebuilding \& Development}} \bibinfo{volume}{17}, \bibinfo{number}{2} (\bibinfo{year}{2022}), \bibinfo{pages}{239--243}.
\newblock
\urldef\tempurl%
\url{https://doi.org/10.1177/15423166221102757}
\showDOI{\tempurl}
\showeprint{https://doi.org/10.1177/15423166221102757}


\bibitem[Argyle et~al\mbox{.}(2023)]%
        {argyle2023leveraging}
\bibfield{author}{\bibinfo{person}{Lisa~P Argyle}, \bibinfo{person}{Christopher~A Bail}, \bibinfo{person}{Ethan~C Busby}, \bibinfo{person}{Joshua~R Gubler}, \bibinfo{person}{Thomas Howe}, \bibinfo{person}{Christopher Rytting}, \bibinfo{person}{Taylor Sorensen}, {and} \bibinfo{person}{David Wingate}.} \bibinfo{year}{2023}\natexlab{}.
\newblock \showarticletitle{{Leveraging AI for Democratic Discourse: Chat Interventions Can Improve Online Political Conversations at Scale}}.
\newblock \bibinfo{journal}{\emph{Proceedings of the National Academy of Sciences}} \bibinfo{volume}{120}, \bibinfo{number}{41} (\bibinfo{year}{2023}), \bibinfo{pages}{e2311627120}.
\newblock


\bibitem[Bakker et~al\mbox{.}(2022)]%
        {bakker2022fine}
\bibfield{author}{\bibinfo{person}{Michiel Bakker}, \bibinfo{person}{Martin Chadwick}, \bibinfo{person}{Hannah Sheahan}, \bibinfo{person}{Michael Tessler}, \bibinfo{person}{Lucy Campbell-Gillingham}, \bibinfo{person}{Jan Balaguer}, \bibinfo{person}{Nat McAleese}, \bibinfo{person}{Amelia Glaese}, \bibinfo{person}{John Aslanides}, \bibinfo{person}{Matt Botvinick}, {et~al\mbox{.}}} \bibinfo{year}{2022}\natexlab{}.
\newblock \showarticletitle{{Fine-tuning Language Models to Find Agreement Among Humans with Diverse Preferences}}.
\newblock \bibinfo{journal}{\emph{Advances in Neural Information Processing Systems}}  \bibinfo{volume}{35} (\bibinfo{year}{2022}), \bibinfo{pages}{38176--38189}.
\newblock


\bibitem[Bilich et~al\mbox{.}(2023)]%
        {bilich2019faster}
\bibfield{author}{\bibinfo{person}{Jordan Bilich}, \bibinfo{person}{Michael Varga}, \bibinfo{person}{Daanish Masood}, {and} \bibinfo{person}{Andrew Konya}.} \bibinfo{year}{2023}\natexlab{}.
\newblock \bibinfo{title}{{Faster Peace via Inclusivity: An Efficient Paradigm to Understand Populations in Conflict Zones}}.
\newblock
\newblock
\showeprint[arxiv]{2311.00816}~[cs.CY]
\urldef\tempurl%
\url{https://arxiv.org/abs/2311.00816}
\showURL{%
\tempurl}


\bibitem[Brandt et~al\mbox{.}(2016)]%
        {brandt2016handbook}
\bibfield{author}{\bibinfo{person}{Felix Brandt}, \bibinfo{person}{Vincent Conitzer}, \bibinfo{person}{Ulle Endriss}, \bibinfo{person}{J\'{e}r\^{o}me Lang}, {and} \bibinfo{person}{Ariel~D. Procaccia}.} \bibinfo{year}{2016}\natexlab{}.
\newblock \bibinfo{booktitle}{\emph{{Handbook of Computational Social Choice}} (\bibinfo{edition}{1st} ed.)}.
\newblock \bibinfo{publisher}{Cambridge University Press}, \bibinfo{address}{USA}.
\newblock
\showISBNx{1107060435}


\bibitem[Bravo-Hermsdorff et~al\mbox{.}(2022)]%
        {bravohermsdorff2022statistical}
\bibfield{author}{\bibinfo{person}{Gecia Bravo-Hermsdorff}, \bibinfo{person}{Robert Busa-Fekete}, \bibinfo{person}{Lee~M. Gunderson}, \bibinfo{person}{Andrés~Munõz Medina}, {and} \bibinfo{person}{Umar Syed}.} \bibinfo{year}{2022}\natexlab{}.
\newblock \bibinfo{title}{{Statistical Anonymity: Quantifying Reidentification Risks without Reidentifying Users}}.
\newblock
\newblock
\showeprint[arxiv]{2201.12306}~[cs.DS]
\urldef\tempurl%
\url{https://arxiv.org/abs/2201.12306}
\showURL{%
\tempurl}


\bibitem[{British Columbia Citizens' Assembly on Electoral Reform}(2004)]%
        {BCAssembly2004}
\bibfield{author}{\bibinfo{person}{{British Columbia Citizens' Assembly on Electoral Reform}}.} \bibinfo{year}{2004}\natexlab{}.
\newblock \bibinfo{title}{{Making Every Vote Count: The Case for Electoral Reform in British Columbia}}.
\newblock
\newblock
\urldef\tempurl%
\url{http://citizensassembly.arts.ubc.ca/resources/final_report.pdf}
\showURL{%
\tempurl}


\bibitem[{B{\"u}rgerBegehren Klimaschutz}(2020)]%
        {BuergerRat2020}
\bibfield{author}{\bibinfo{person}{{B{\"u}rgerBegehren Klimaschutz}}.} \bibinfo{year}{2020}\natexlab{}.
\newblock \bibinfo{title}{{Climate Assembly Adopts Recommendations}}.
\newblock \bibinfo{howpublished}{Press Release, B{\"u}rgerRat Democracy}.
\newblock
\urldef\tempurl%
\url{https://www.buergerrat.de/en/news/climate-assembly-adopts-recommendations/}
\showURL{%
\tempurl}


\bibitem[Burton et~al\mbox{.}(2024)]%
        {burton2024large}
\bibfield{author}{\bibinfo{person}{Jason~W. Burton}, \bibinfo{person}{Ezequiel Lopez-Lopez}, \bibinfo{person}{Shahar Hechtlinger}, \bibinfo{person}{Zoe Rahwan}, \bibinfo{person}{Samuel Aeschbach}, \bibinfo{person}{Michiel~A. Bakker}, \bibinfo{person}{Joshua~A. Becker}, \bibinfo{person}{Aleks Berditchevskaia}, \bibinfo{person}{Julian Berger}, \bibinfo{person}{Levin Brinkmann}, \bibinfo{person}{Lucie Flek}, \bibinfo{person}{Stefan~M. Herzog}, \bibinfo{person}{Saffron Huang}, \bibinfo{person}{Sayash Kapoor}, \bibinfo{person}{Arvind Narayanan}, \bibinfo{person}{Anne-Marie Nussberger}, \bibinfo{person}{Taha Yasseri}, \bibinfo{person}{Pietro Nickl}, \bibinfo{person}{Abdullah Almaatouq}, \bibinfo{person}{Ulrike Hahn}, \bibinfo{person}{Ralf H. J.~M. Kurvers}, \bibinfo{person}{Susan Leavy}, \bibinfo{person}{Iyad Rahwan}, \bibinfo{person}{Divya Siddarth}, \bibinfo{person}{Alice Siu}, \bibinfo{person}{Anita~W. Woolley}, \bibinfo{person}{Dirk~U. Wulff}, {and} \bibinfo{person}{Ralph Hertwig}.}
  \bibinfo{year}{2024}\natexlab{}.
\newblock \showarticletitle{{How large language models can reshape collective intelligence}}.
\newblock \bibinfo{journal}{\emph{Nature Human Behaviour}} \bibinfo{volume}{8}, \bibinfo{number}{9} (\bibinfo{date}{01 Sep} \bibinfo{year}{2024}), \bibinfo{pages}{1643--1655}.
\newblock
\showISSN{2397-3374}
\urldef\tempurl%
\url{https://doi.org/10.1038/s41562-024-01959-9}
\showDOI{\tempurl}


\bibitem[De et~al\mbox{.}(2025)]%
        {de2025supernotes}
\bibfield{author}{\bibinfo{person}{Soham De}, \bibinfo{person}{Jay Baxter}, \bibinfo{person}{Michiel Bakker}, {and} \bibinfo{person}{Martin Saveski}.} \bibinfo{year}{2025}\natexlab{}.
\newblock \showarticletitle{{Supernotes: Driving Consensus in Crowd-Sourced Fact-Checking}}. In \bibinfo{booktitle}{\emph{Proceedings of the ACM Web Conference 2025 (WWW '25)}} (Sydney, NSW, Australia). \bibinfo{publisher}{ACM}, \bibinfo{address}{New York, NY, USA}, \bibinfo{pages}{11}.
\newblock
\urldef\tempurl%
\url{https://doi.org/10.1145/3696410.3714934}
\showDOI{\tempurl}


\bibitem[Dillard(2013)]%
        {dillard2013}
\bibfield{author}{\bibinfo{person}{Kara~N. Dillard}.} \bibinfo{year}{2013}\natexlab{}.
\newblock \showarticletitle{{Envisioning the Role of Facilitation in Public Deliberation}}.
\newblock \bibinfo{journal}{\emph{Journal of Applied Communication Research}} \bibinfo{volume}{41}, \bibinfo{number}{3} (\bibinfo{year}{2013}), \bibinfo{pages}{217--235}.
\newblock
\urldef\tempurl%
\url{https://doi.org/10.1080/00909882.2013.826813}
\showDOI{\tempurl}


\bibitem[Farrell et~al\mbox{.}(2015)]%
        {Farrell2015MarriageEquality}
\bibfield{author}{\bibinfo{person}{David Farrell}, \bibinfo{person}{Clodagh Harris}, {and} \bibinfo{person}{Jane Suiter}.} \bibinfo{year}{2015}\natexlab{}.
\newblock \showarticletitle{{The Irish Vote for Marriage Equality Started at a Constitutional Convention}}.
\newblock \bibinfo{journal}{\emph{The Washington Post (Monkey Cage)}} (\bibinfo{year}{2015}).
\newblock
\urldef\tempurl%
\url{https://www.washingtonpost.com/news/monkey-cage/wp/2015/06/05/the-irish-vote-for-marriage-equality-started-at-a-constitutional-convention/}
\showURL{%
\tempurl}


\bibitem[Fish et~al\mbox{.}(2025)]%
        {fish2023generative}
\bibfield{author}{\bibinfo{person}{Sara Fish}, \bibinfo{person}{Paul Gölz}, \bibinfo{person}{David~C. Parkes}, \bibinfo{person}{Ariel~D. Procaccia}, \bibinfo{person}{Gili Rusak}, \bibinfo{person}{Itai Shapira}, {and} \bibinfo{person}{Manuel Wüthrich}.} \bibinfo{year}{2025}\natexlab{}.
\newblock \bibinfo{title}{{Generative Social Choice}}.
\newblock
\newblock
\showeprint[arxiv]{2309.01291}~[cs.GT]
\urldef\tempurl%
\url{https://arxiv.org/abs/2309.01291}
\showURL{%
\tempurl}


\bibitem[Fraenkel and Grofman(2014)]%
        {fraenkel2014borda}
\bibfield{author}{\bibinfo{person}{Jon Fraenkel} {and} \bibinfo{person}{Bernard Grofman}.} \bibinfo{year}{2014}\natexlab{}.
\newblock \showarticletitle{{The Borda Count and its Real-World Alternatives: Comparing Scoring Rules in Nauru and Slovenia}}.
\newblock \bibinfo{journal}{\emph{Australian Journal of Political Science}} \bibinfo{volume}{49}, \bibinfo{number}{2} (\bibinfo{year}{2014}), \bibinfo{pages}{186--205}.
\newblock


\bibitem[Gabbatiss(2020)]%
        {CarbonBrief2020}
\bibfield{author}{\bibinfo{person}{Josh Gabbatiss}.} \bibinfo{year}{2020}\natexlab{}.
\newblock \showarticletitle{{Q\&A: How the `Climate Assembly' Says the UK Should Reach Net-Zero}}.
\newblock \bibinfo{journal}{\emph{Carbon Brief}} (\bibinfo{year}{2020}).
\newblock
\urldef\tempurl%
\url{https://www.carbonbrief.org/qa-how-the-climate-assembly-says-the-uk-should-reach-net-zero/}
\showURL{%
\tempurl}


\bibitem[Gaffney(2022)]%
        {gaffney2022impartiality}
\bibfield{author}{\bibinfo{person}{Imelda Gaffney}.} \bibinfo{year}{2022}\natexlab{}.
\newblock \showarticletitle{{Impartiality and Neutrality in Mediation}}.
\newblock \bibinfo{journal}{\emph{Journal of Mediation and Applied Conflict Analysis}} \bibinfo{volume}{8}, \bibinfo{number}{1} (\bibinfo{year}{2022}), \bibinfo{pages}{59--71}.
\newblock


\bibitem[{Global Assembly}(2022)]%
        {GlobalAssembly2022}
\bibfield{author}{\bibinfo{person}{{Global Assembly}}.} \bibinfo{year}{2022}\natexlab{}.
\newblock \bibinfo{title}{{Report of the 2021 Global Assembly on the Climate and Ecological Crisis}}.
\newblock
\newblock
\urldef\tempurl%
\url{https://globalassembly.org/resources/downloads/GlobalAssembly2021-FullReport.pdf}
\showURL{%
\tempurl}


\bibitem[Goldberg et~al\mbox{.}(2024)]%
        {goldberg2024ai}
\bibfield{author}{\bibinfo{person}{Beth Goldberg}, \bibinfo{person}{Diana Acosta-Navas}, \bibinfo{person}{Michiel Bakker}, \bibinfo{person}{Ian Beacock}, \bibinfo{person}{Matt Botvinick}, \bibinfo{person}{Prateek Buch}, \bibinfo{person}{Renée DiResta}, \bibinfo{person}{Nandika Donthi}, \bibinfo{person}{Nathanael Fast}, \bibinfo{person}{Ravi Iyer}, \bibinfo{person}{Zaria Jalan}, \bibinfo{person}{Andrew Konya}, \bibinfo{person}{Grace~Kwak Danciu}, \bibinfo{person}{Hélène Landemore}, \bibinfo{person}{Alice Marwick}, \bibinfo{person}{Carl Miller}, \bibinfo{person}{Aviv Ovadya}, \bibinfo{person}{Emily Saltz}, \bibinfo{person}{Lisa Schirch}, \bibinfo{person}{Dalit Shalom}, \bibinfo{person}{Divya Siddarth}, \bibinfo{person}{Felix Sieker}, \bibinfo{person}{Christopher Small}, \bibinfo{person}{Jonathan Stray}, \bibinfo{person}{Audrey Tang}, \bibinfo{person}{Michael~Henry Tessler}, {and} \bibinfo{person}{Amy Zhang}.} \bibinfo{year}{2024}\natexlab{}.
\newblock \bibinfo{title}{{AI and the Future of Digital Public Squares}}.
\newblock
\newblock
\showeprint[arxiv]{2412.09988}~[cs.CY]
\urldef\tempurl%
\url{https://arxiv.org/abs/2412.09988}
\showURL{%
\tempurl}


\bibitem[Involve(2018)]%
        {Involve2018}
\bibfield{author}{\bibinfo{person}{Involve}.} \bibinfo{year}{2018}\natexlab{}.
\newblock \bibinfo{title}{{The Citizens' Assembly Behind The Irish Abortion Referendum}}.
\newblock
\newblock
\urldef\tempurl%
\url{https://involve.org.uk/news-opinion/opinion/citizens-assembly-behind-irish-abortion-referendum}
\showURL{%
\tempurl}


\bibitem[Konya et~al\mbox{.}(2024)]%
        {konya2024chain}
\bibfield{author}{\bibinfo{person}{Andrew Konya}, \bibinfo{person}{Aviv Ovadya}, \bibinfo{person}{Kevin Feng}, \bibinfo{person}{Quan~Ze Chen}, \bibinfo{person}{Lisa Schirch}, \bibinfo{person}{Colin Irwin}, {and} \bibinfo{person}{Amy~X. Zhang}.} \bibinfo{year}{2024}\natexlab{}.
\newblock \bibinfo{title}{{Chain of Alignment: Integrating Public Will with Expert Intelligence for Language Model Alignment}}.
\newblock
\newblock
\showeprint[arxiv]{2411.10534}~[cs.HC]
\urldef\tempurl%
\url{https://arxiv.org/abs/2411.10534}
\showURL{%
\tempurl}


\bibitem[Konya et~al\mbox{.}(2022)]%
        {konya2022elicitation}
\bibfield{author}{\bibinfo{person}{Andrew Konya}, \bibinfo{person}{Yeping~Lina Qiu}, \bibinfo{person}{Michael~P Varga}, {and} \bibinfo{person}{Aviv Ovadya}.} \bibinfo{year}{2022}\natexlab{}.
\newblock \showarticletitle{{Elicitation Inference Optimization for Multi-Principal-Agent Alignment}}. In \bibinfo{booktitle}{\emph{NeurIPS 2022 Foundation Models for Decision Making Workshop}}.
\newblock
\urldef\tempurl%
\url{https://openreview.net/forum?id=tkxnRPkb_H}
\showURL{%
\tempurl}


\bibitem[Konya et~al\mbox{.}(2023)]%
        {konya2023dem}
\bibfield{author}{\bibinfo{person}{Andrew Konya}, \bibinfo{person}{Lisa Schirch}, \bibinfo{person}{Colin Irwin}, {and} \bibinfo{person}{Aviv Ovadya}.} \bibinfo{year}{2023}\natexlab{}.
\newblock \bibinfo{title}{{Democratic Policy Development using Collective Dialogues and AI}}.
\newblock
\newblock
\showeprint[arxiv]{2311.02242}~[cs.CY]
\urldef\tempurl%
\url{https://arxiv.org/abs/2311.02242}
\showURL{%
\tempurl}


\bibitem[Lindskold(1978)]%
        {lindskold1978trust}
\bibfield{author}{\bibinfo{person}{Svenn Lindskold}.} \bibinfo{year}{1978}\natexlab{}.
\newblock \showarticletitle{{Trust Development, the GRIT Proposal, and the Effects of Conciliatory Acts on Conflict and Cooperation}}.
\newblock \bibinfo{journal}{\emph{Psychological Bulletin}} \bibinfo{volume}{85}, \bibinfo{number}{4} (\bibinfo{year}{1978}), \bibinfo{pages}{772}.
\newblock


\bibitem[Masood~Alavi et~al\mbox{.}(2022)]%
        {masood2022using}
\bibfield{author}{\bibinfo{person}{Daanish Masood~Alavi}, \bibinfo{person}{Martin W{\"a}hlisch}, \bibinfo{person}{Colin Irwin}, {and} \bibinfo{person}{Andrew Konya}.} \bibinfo{year}{2022}\natexlab{}.
\newblock \showarticletitle{{Using Artificial Intelligence for Peacebuilding}}.
\newblock \bibinfo{journal}{\emph{Journal of Peacebuilding \& Development}} \bibinfo{volume}{17}, \bibinfo{number}{2} (\bibinfo{year}{2022}), \bibinfo{pages}{239--243}.
\newblock


\bibitem[Montville(1991)]%
        {montville1991track}
\bibfield{author}{\bibinfo{person}{Joseph Montville}.} \bibinfo{year}{1991}\natexlab{}.
\newblock \showarticletitle{{Track Two Diplomacy: The Arrow and the Olive Branch}}.
\newblock In \bibinfo{booktitle}{\emph{The Psychodynamics of International Relations}}, \bibfield{editor}{\bibinfo{person}{Vamik~D. Volkan}, \bibinfo{person}{Demetrios~A. Julius}, {and} \bibinfo{person}{Joseph~V. Montville}} (Eds.). Vol.~\bibinfo{volume}{2}. \bibinfo{publisher}{Lexington Books}, \bibinfo{address}{Lexington, MA}, \bibinfo{pages}{161--175}.
\newblock


\bibitem[Moore(2014)]%
        {moore2014}
\bibfield{author}{\bibinfo{person}{Christopher~W. Moore}.} \bibinfo{year}{2014}\natexlab{}.
\newblock \bibinfo{booktitle}{\emph{{The Mediation Process: Practical Strategies for Resolving Conflict}}}.
\newblock \bibinfo{publisher}{Jossey-Bass}, \bibinfo{address}{{San Francisco, CA, USA}}.
\newblock
\showISBNx{978-1-118-30430-3}


\bibitem[Ovadya(2022)]%
        {ovadya2022bridging}
\bibfield{author}{\bibinfo{person}{Aviv Ovadya}.} \bibinfo{year}{2022}\natexlab{}.
\newblock \bibinfo{booktitle}{\emph{{Bridging-Based Ranking}}}.
\newblock \bibinfo{type}{{T}echnical {R}eport}. \bibinfo{institution}{Belfer Center for Science and International Affairs, Harvard Kennedy School}.
\newblock
\urldef\tempurl%
\url{https://www.belfercenter.org/publication/bridging-based-ranking}
\showURL{%
\tempurl}


\bibitem[Ovadya(2023)]%
        {ovadya2023generative}
\bibfield{author}{\bibinfo{person}{Aviv Ovadya}.} \bibinfo{year}{2023}\natexlab{}.
\newblock \bibinfo{title}{{'Generative CI' through Collective Response Systems}}.
\newblock
\newblock
\showeprint[arxiv]{2302.00672}~[cs.HC]
\urldef\tempurl%
\url{https://arxiv.org/abs/2302.00672}
\showURL{%
\tempurl}


\bibitem[Ovadya and Thorburn(2023)]%
        {ovadya2023}
\bibfield{author}{\bibinfo{person}{Aviv Ovadya} {and} \bibinfo{person}{Luke Thorburn}.} \bibinfo{year}{2023}\natexlab{}.
\newblock \bibinfo{booktitle}{\emph{{Bridging Systems: Open Problems for Countering Destructive Divisiveness across Ranking, Recommenders, and Governance}}}.
\newblock \bibinfo{type}{{T}echnical {R}eport}. \bibinfo{institution}{Knight First Amendment Institute}.
\newblock
\urldef\tempurl%
\url{https://knightcolumbia.org/content/bridging-systems}
\showURL{%
\tempurl}


\bibitem[Schirch and Campt(2015)]%
        {schirch2015}
\bibfield{author}{\bibinfo{person}{Lisa Schirch} {and} \bibinfo{person}{David Campt}.} \bibinfo{year}{2015}\natexlab{}.
\newblock \bibinfo{booktitle}{\emph{{The Little Book of Dialogue for Difficult Subjects: A Practical, Hands-On Guide}}}.
\newblock \bibinfo{publisher}{Simon and Schuster}, \bibinfo{address}{{New York, USA}}.
\newblock
\showISBNx{978-1-68099-030-0}


\bibitem[Small(2015)]%
        {polisMathCommit8394f1f}
\bibfield{author}{\bibinfo{person}{Christopher Small}.} \bibinfo{year}{2015}\natexlab{}.
\newblock \bibinfo{title}{{Commit 8394f1f}}.
\newblock
\newblock
\urldef\tempurl%
\url{https://github.com/pol-is/polisMath/commit/8394f1fa75fd64d0f9eee9335f74d2fa0731f882}
\showURL{%
\tempurl}


\bibitem[Small et~al\mbox{.}(2021)]%
        {small2021polis}
\bibfield{author}{\bibinfo{person}{Christopher Small}, \bibinfo{person}{Michael Bjorkegren}, \bibinfo{person}{Timo Erkkil{\"a}}, \bibinfo{person}{Lynette Shaw}, {and} \bibinfo{person}{Colin Megill}.} \bibinfo{year}{2021}\natexlab{}.
\newblock \showarticletitle{{Polis: Scaling Deliberation by Mapping High Dimensional Opinion Spaces}}.
\newblock \bibinfo{journal}{\emph{Recerca: Revista de Pensament i An{\`a}lisi}} \bibinfo{volume}{26}, \bibinfo{number}{2} (\bibinfo{year}{2021}).
\newblock


\bibitem[Small et~al\mbox{.}(2023)]%
        {small2023opportunities}
\bibfield{author}{\bibinfo{person}{Christopher~T. Small}, \bibinfo{person}{Ivan Vendrov}, \bibinfo{person}{Esin Durmus}, \bibinfo{person}{Hadjar Homaei}, \bibinfo{person}{Elizabeth Barry}, \bibinfo{person}{Julien Cornebise}, \bibinfo{person}{Ted Suzman}, \bibinfo{person}{Deep Ganguli}, {and} \bibinfo{person}{Colin Megill}.} \bibinfo{year}{2023}\natexlab{}.
\newblock \bibinfo{title}{{Opportunities and Risks of LLMs for Scalable Deliberation with Polis}}.
\newblock
\newblock
\showeprint[arxiv]{2306.11932}~[cs.SI]
\urldef\tempurl%
\url{https://arxiv.org/abs/2306.11932}
\showURL{%
\tempurl}


\bibitem[Tessler et~al\mbox{.}(2024)]%
        {tessler2024ai}
\bibfield{author}{\bibinfo{person}{Michael~Henry Tessler}, \bibinfo{person}{Michiel~A Bakker}, \bibinfo{person}{Daniel Jarrett}, \bibinfo{person}{Hannah Sheahan}, \bibinfo{person}{Martin~J Chadwick}, \bibinfo{person}{Raphael Koster}, \bibinfo{person}{Georgina Evans}, \bibinfo{person}{Lucy Campbell-Gillingham}, \bibinfo{person}{Tantum Collins}, \bibinfo{person}{David~C Parkes}, {et~al\mbox{.}}} \bibinfo{year}{2024}\natexlab{}.
\newblock \showarticletitle{{AI Can Help Humans Find Common Ground in Democratic Deliberation}}.
\newblock \bibinfo{journal}{\emph{Science}} \bibinfo{volume}{386}, \bibinfo{number}{6719} (\bibinfo{year}{2024}), \bibinfo{pages}{eadq2852}.
\newblock


\bibitem[{United Nations}(2008)]%
        {un_peacekeeping_2008}
\bibfield{author}{\bibinfo{person}{{United Nations}}.} \bibinfo{year}{2008}\natexlab{}.
\newblock \bibinfo{title}{{United Nations Peacekeeping Operations: Principles and Guidelines}}.
\newblock , \bibinfo{numpages}{31--35}~pages.
\newblock
\urldef\tempurl%
\url{https://peacekeeping.un.org/sites/default/files/capstone_eng_0.pdf}
\showURL{%
\tempurl}


\bibitem[{United Nations DPPA Innovation}(2022a)]%
        {UNBahrain}
\bibfield{author}{\bibinfo{person}{{United Nations DPPA Innovation}}.} \bibinfo{year}{2022}\natexlab{a}.
\newblock \bibinfo{title}{{Project: Berlanti}}.
\newblock
\newblock
\urldef\tempurl%
\url{https://futuringpeace.org/project/berlanti}
\showURL{%
\tempurl}


\bibitem[{United Nations DPPA Innovation}(2022b)]%
        {UNBolivia}
\bibfield{author}{\bibinfo{person}{{United Nations DPPA Innovation}}.} \bibinfo{year}{2022}\natexlab{b}.
\newblock \bibinfo{title}{{Project: Digital Dialogues Bolivia}}.
\newblock
\newblock
\urldef\tempurl%
\url{https://futuringpeace.org/project/project-digital-dialogues-haiti}
\showURL{%
\tempurl}


\bibitem[{United Nations DPPA Innovation}(2022c)]%
        {UNHati}
\bibfield{author}{\bibinfo{person}{{United Nations DPPA Innovation}}.} \bibinfo{year}{2022}\natexlab{c}.
\newblock \bibinfo{title}{{Project: Digital Dialogues Haiti}}.
\newblock
\newblock
\urldef\tempurl%
\url{https://futuringpeace.org/project/project-digital-dialogues-haiti}
\showURL{%
\tempurl}


\bibitem[{United Nations DPPA Innovation}(2022d)]%
        {UNIraq}
\bibfield{author}{\bibinfo{person}{{United Nations DPPA Innovation}}.} \bibinfo{year}{2022}\natexlab{d}.
\newblock \bibinfo{title}{{Project: Digital Dialogues Iraq}}.
\newblock
\newblock
\urldef\tempurl%
\url{https://futuringpeace.org/project/project-digital-dialogues-iraq}
\showURL{%
\tempurl}


\bibitem[{United Nations DPPA Innovation}(2022e)]%
        {UNLebanon}
\bibfield{author}{\bibinfo{person}{{United Nations DPPA Innovation}}.} \bibinfo{year}{2022}\natexlab{e}.
\newblock \bibinfo{title}{{Project: Digital Dialogues Lebanon}}.
\newblock
\newblock
\urldef\tempurl%
\url{https://futuringpeace.org/project/project-digital-dialogues-lebanon}
\showURL{%
\tempurl}


\bibitem[{United Nations DPPA Innovation}(2022f)]%
        {UNLibya}
\bibfield{author}{\bibinfo{person}{{United Nations DPPA Innovation}}.} \bibinfo{year}{2022}\natexlab{f}.
\newblock \bibinfo{title}{{Project: Digital Dialogues Libya}}.
\newblock
\newblock
\urldef\tempurl%
\url{https://futuringpeace.org/project/project-digital-dialogues-libya}
\showURL{%
\tempurl}


\bibitem[{United Nations DPPA Innovation}(2022g)]%
        {futuringPeaceDigitalDialoguesYemen}
\bibfield{author}{\bibinfo{person}{{United Nations DPPA Innovation}}.} \bibinfo{year}{2022}\natexlab{g}.
\newblock \bibinfo{title}{{Project: Digital Dialogues Yemen}}.
\newblock
\newblock
\urldef\tempurl%
\url{https://futuringpeace.org/project/project-digital-dialogues-yemen}
\showURL{%
\tempurl}


\bibitem[{United Nations DPPA Innovation}(2022h)]%
        {UNBolivia2}
\bibfield{author}{\bibinfo{person}{{United Nations DPPA Innovation}}.} \bibinfo{year}{2022}\natexlab{h}.
\newblock \bibinfo{title}{{Project: Sergio's Conversa}}.
\newblock
\newblock
\urldef\tempurl%
\url{https://futuringpeace.org/project/sergios-conversa}
\showURL{%
\tempurl}


\bibitem[{van Gelder}(2012)]%
        {van-gelder2012}
\bibfield{author}{\bibinfo{person}{Tim {van Gelder}}.} \bibinfo{year}{2012}\natexlab{}.
\newblock \showarticletitle{{Cultivating Deliberation for Democracy}}.
\newblock \bibinfo{journal}{\emph{Journal of Deliberative Democracy}} \bibinfo{volume}{8}, \bibinfo{number}{1} (\bibinfo{year}{2012}).
\newblock
\urldef\tempurl%
\url{https://doi.org/10.16997/jdd.134}
\showDOI{\tempurl}


\bibitem[Wojcik et~al\mbox{.}(2022)]%
        {wojcik2022bird}
\bibfield{author}{\bibinfo{person}{Stefan Wojcik}, \bibinfo{person}{Sophie Hilgard}, \bibinfo{person}{Nick Judd}, \bibinfo{person}{Delia Mocanu}, \bibinfo{person}{Stephen Ragain}, \bibinfo{person}{M.~B.~Fallin Hunzaker}, \bibinfo{person}{Keith Coleman}, {and} \bibinfo{person}{Jay Baxter}.} \bibinfo{year}{2022}\natexlab{}.
\newblock \bibinfo{title}{{Birdwatch: Crowd Wisdom and Bridging Algorithms can Inform Understanding and Reduce the Spread of Misinformation}}.
\newblock
\newblock
\showeprint[arxiv]{2210.15723}~[cs.SI]
\urldef\tempurl%
\url{https://arxiv.org/abs/2210.15723}
\showURL{%
\tempurl}


\end{thebibliography}

\clearpage
\appendix

\section{Joint Letter}
\label{appendix:letter}

    {\itshape ``We write as a united body of Israeli and Palestinian peacebuilders, compelled by the urgency of our shared pain and galvanized with the hope we still hold for the future. We call for the following:
        
    \begin{itemize}
        \item An immediate ceasefire and release of all hostages
        \item The creation of an international fund to build the foundations for a comprehensive long-term solution
        \item The establishment of a globalized coalition to assist Gaza's reconstruction 
        \item The inclusion of civil society and NGOs in all peacebuilding and diplomatic discussions
        \item The recognition of both sides' trauma
    \end{itemize}
    
    Acknowledging the trauma and pain experienced by both sides is essential for the healing of our societies and the achievement of sustainable peace. Help us break the cycle of violence so the next generation may flourish. The world watches, and history will judge. Now is the time to act.''}

\section{Further Background}
\subsection{Collective Dialogue Dynamics}
\label{appendix:collective}

    During a live collective dialogue process, all participation is simultaneous and each collective response process takes a few minutes. When each response process completes, each participant sees the fraction of all participants agreeing with their response along with a representative subset of responses from the group. The moderator sees live results which they can use to decide if they need to pivot from the pre-programmed discussion guide to ask additional questions on an issue. After a live dialogue is complete it can be left open for asynchronous participation so those who could not make the live event can still participate. Collective dialogues can also be run asynchronously from the start if the logistics of a live dialogue are not feasible for the situation.

\section{Further Methodological Details}

\subsection{Transparency about the use of AI}
\label{appendix:transparency}

    In the invitations and public communications, participants were told ``AI'' was being used, e.g., ``we are using an AI-based tool to hold a series of large scale `collective dialogues' among Israelis and Palestinians.'' \ref{appendix:invite} Some participants attended optional office hour/Q\&A calls, in which we discussed how LLMs specifically were being used.

    Most prospective participants responded positively. Some were skeptical of the use of AI and wanted to understand the details of how it was being used further. For those, we held a few ‘office hour’ sessions where we walked anyone interested or concerned through the details of how we used AI. Most questions were around elicitation inference \& how well LLMs could handle Hebrew/Arabic. We typically walked them through the overall process, explained the micro-experiments for the LLM pipeline related to Hebrew/Arabic, and shared links to the elicitation inference papers. 

    It’s unclear exactly how awareness of the use of LLMs shaped the consensus-building process, but we think that talking to a ``computer'' made it easier for people to express themselves freely. For some, LLMs were seen as more impartial. For others, there was persistent skepticism.

\section{Further Discussion}
\label{appendix:discussion}

\subsection{Ethical Considerations}

    Undertaking this work this in the context of active conflict raises many valid ethical concerns. We summarize some of those we considered below, along with steps taken to address them. In general, when making decisions about delaying or potentially not running the process, we weighed the potential downsides and upsides of moving forward against the ``null hypothesis'' of the status quo.
    \begin{itemize}
        
        \item \textbf{Backfire}: There was a risk that conducting the cross-border dialogue in the joint phase could backfire in various ways (e.g., if dialogue broke down), and inflame the division this effort sought to bridge. We mitigated this risk by starting with a uninational phase designed to help build trust and set the conditions for a productive cross-border dialogue.
        
        \item \textbf{Bias}: Any bias in the process could harm the legitimacy of the results it produced.  There are many ways bias might have entered the process, some of which we list below, along with the steps taken to mitigate them.
        \begin{itemize}
            
            \item The use of imperfect AI and ML algorithms throughout the process risks injecting errors or biases that misrepresent participants. We mitigated this risk via humans in the loop: experts reviewed and refined all outputs downstream of AI/ML, and final votes among participants were conducted to validate all  points of common ground. To minimize bias from errors in machine translation or LLM processing, we made LLM pipeline choices using micro experiments that enabled native Hebrew and Palestinian Arabic speakers decide what would be used for their language.
            
            \item There was a risk that the human experts ``in the loop'' could themselves induce unfair bias in the process or results. We mitigated this risk by having joint reviews involving both Israelis and Palestinian reviewers at the key points in the process where such biases could be introduced. This included not just process decisions made by the ALLMEP team, but decisions made by the largely US and UK-based team that performed the technical data processing steps in the process. We ensured that there was both Israeli and Palestinian representation on this technical team.
            
            \item For the process to be fair, each ``side'' should have equal influence on the results. But it would be anti-inclusive to not allow certain participants to join in order to ensure equal numbers of participants from all sides. We used equal power metrics to ensure each side had equal influence without requiring equal numbers of participants.
        
            \item To avoid participant self-selection bias on the basis of language or bias due to language difficulties during the process, the approach was designed so that all participants could participate in their primary language.
            
        \end{itemize}
        
        \item \textbf{Malicious use}: There is a risk that results from the process could be used maliciously to disrupt ongoing peacebuilding efforts; i.e., by sewing further division. We thus only share a limited subset of results that manifest common ground or are directly relevant to understanding the approach. We do not share any side-specific quantitative results, and only include overall metrics that are equal power. 

        \item \textbf{Anonymity}: Sharing politically sensitive opinions can potentially present a risk to participants if their identify is known; and force a trade-off between honesty and safety. We thus choose to support all participation to take place anonymously.
     
    \end{itemize}

\subsection{Limitations}

    Several limitations of this case study deserve mention:
    \begin{itemize}
        
        \item \textbf{Lack of controlled experimentation}: Due to the real-world and time-sensitive nature of this work, extensive controlled experimentation was not feasible. This limits the ability to draw strong causal claims or to isolate the impact of specific methodological components.
        
        \item \textbf{Participant selection bias}: The study participants were self-selected peacebuilders and thus not be representative of the broader Israeli and Palestinian populations. It remains unclear if the approach will generalize of to a wider public who may hold more entrenched or divergent viewpoints.
        
        \item \textbf{Context-specificity}: The results are specific to the Israeli-Palestinian context and may not be directly transferable to other conflict zones or social divisions, due to unique political, cultural and historical dynamics.
        
        \item \textbf{Language nuances}: While significant efforts were made to accommodate multiple languages, the use of machine translation and LLMs could introduce subtle errors and ambiguities. These translations --- particularly those of participant-contributed statements that were voted on during the live dialogues and were not subject to human review --- may not have always perfectly captured the intended meaning or cultural nuances of the original statements.
        
        \item \textbf{Technological dependency}: The reliance on online collective dialogue and LLMs tools may introduce a selection bias based on technological literacy and access. The success of this type of process is contingent on reliable internet access, which is not (yet) universally available.
    
    \end{itemize}

\subsection{Future Directions}

    We see a few key challenges to be addressed in future work:
    \begin{itemize}
    
        \item \textbf{Concrete bridging}: It is typically easier to find diverse agreement on anodyne or abstract statements than on statements that are more specific, prescriptive, or concrete. As a result, bridging techniques based only on agreement signals may have a natural bias towards anodyne statements, making it harder to find common ground that is concrete enough to be useful (for example, as the basis for a peace agreement). Future work is needed to create automated measures of statement ``concreteness'' or ``anodyneness'' that can, for example, be combined with bridging metrics to more effectively identify statements that are both bridging and concrete.
        
        \item \textbf{Going beyond distillation}: AI can be used not just for distilling information, but for generating novel more exploratory proposals. Using an iterative deliberation method, similar to the one proposed by Tessler et. al \cite{tessler2024ai}, an LLM could generate a set of initial collective statements that people then critique and refine. This process could explore novel ideas that emerge through a process of iterative synthesis and refinement, with a focus on balanced perspectives. These combined ideas might not individually represent common ground but, when combined in a balanced way, could reflect a form of reasoned consensus. Such statements could not be derived solely from individual statements with high bridging metrics. It would be essential that the resulting statements from this kind of generative process still be validated with a final vote of agreement among human participants to ensure that they actually represent common ground.
        
        \item \textbf{Legitimate representation}: How do you ensure results are representative of a general population? Ideally, the group of participants would statistically representative of the general population and sampled in a way that had both scientific and procedural legitimacy. However, in practice, obtaining a representative sample for online cross-border dialogues in this region is challenging due to structural, political, and logistical barriers. Restrictions on movement, lack of digital accessibility in certain areas, surveillance concerns, and varying levels of trust in international initiatives all affect willingness and ability to participate. Further, the type of nationally representative online panel that might otherwise serve as a starting point for this type of effort do not exist for areas like the West Bank and Gaza. While establishing such an online panel would be ideal, one short-term work around may be to recruit a diverse (but not statistically representative) set of participants for the online collective dialogues, then use more traditional computer-assisted telephone interviewing and in-person sampling approaches to obtain legitimately representative samples for the final vote, including marginalized populations like rural communities, refugee camps, and those with lower access to digital platforms.
        
    \end{itemize} 

\section{Qualitative Feedback}
\label{appendix:feedback}

\subsection{ALLMEP Retrospective}

    Below are a few specific questions asked to ALLMEP leaders involved in the process. Direct quotes denote their verbatim responses and content not in quotes is context provided by this paper's authors. 

    \begin{itemize}
        \item \textbf{Did you find [the process] helpful?} 
        
        \emph{``Yes. At a time when real dialogue felt impossible post October 7th, this process allowed Palestinians and Israelis to engage without compromising their experiences or red lines.''}
        
        \item \textbf{Do you plan to use the process again?} 
        
        \emph{``Absolutely. The ability to engage anonymously and equitably is critical, especially for Palestinians who face political and social risks when participating in cross-border work.''}
        
        (We are continuing to work with ALLMEP on scaling this process up to larger and more representative samples of the public; of which --- as of writing this --- the first dialogue cycle has already begun.)
        
        \item \textbf{Were there parts of the process that you were especially excited about or skeptical of [before it was run]?}
        
        \emph{``I was mostly excited about the uninational phase. It gave [us] room to define our priorities before entering joint dialogue, which felt empowering and necessary. I was a little skeptical however whether AI could reflect our political realities.''}
        
        Beyond all the trust building aspects discussed throughout the paper, we note that using the well-known Community Notes algorithm helped give the process improved face validity and build trust.
    \end{itemize}

\subsection{Participant Feedback}

\subsubsection{Examples of Positive Feedback}

    \begin{itemize}
        
        \item ``I found it to be very validating and very positive! It gave me hope''
        
        \item ``That was an INCREDIBLE experience''
        
        \item ``I felt that I could share openly, without fear of being judged. It felt much less complicated than sharing directly with colleagues''
        
        \item ``This was an interesting experience, a lot of self-reflection regarding our inner motives for this work''
        
        \item ``I know there is an end goal for this, but seeing each other's anonymous thoughts in this way is valuable in and of itself and perhaps this sort of work could be integrated in various ways as part of more regular activities. The openness of anonymity is important.''
    
    \end{itemize}

\subsubsection{Examples of Negative or Ambivalent Feedback}

    \begin{itemize}
        
        \item ``Interesting activity but time for answers was a bit short''
        
        \item ``Ummm.... I haven't made up my mind about this technology yet....''
        
        \item ``Everything is so fast. It's hard to think. It was possible to give a little more time to read and respond. The idea is cool''
        
        \item ``We need to be able to speak to each other, see each other's faces''
        
        \item ``I am not sure that I was able to express myself completely or clearly also due to technical limitations of time and the translation.''

    \end{itemize}

    \clearpage
    \onecolumn

\section{Recruitment Invitation}
\label{appendix:invite}

    \vspace{5mm}
    \begin{center}
        \fbox{\includegraphics[width=0.9\textwidth]{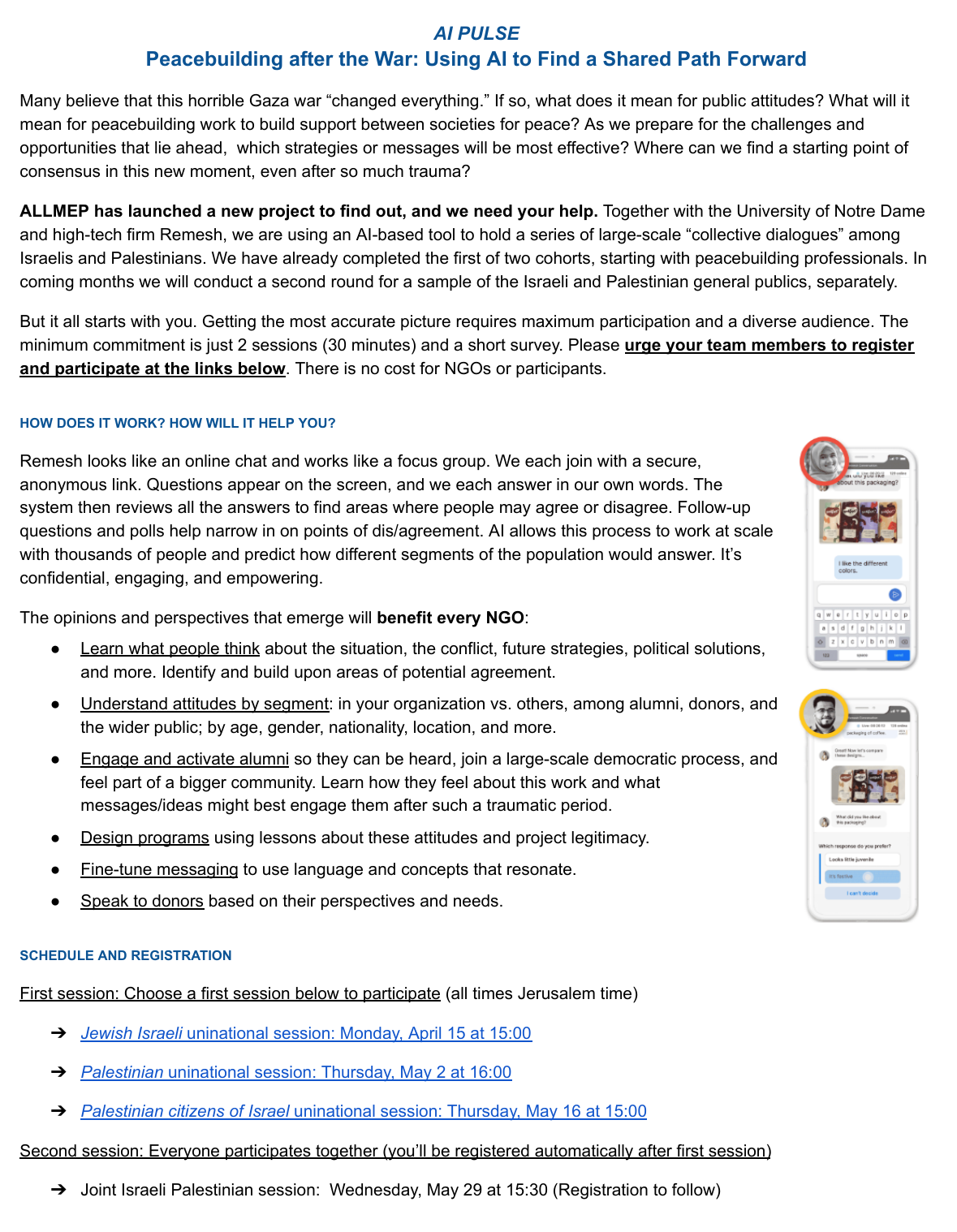}}
    \end{center}
    \clearpage

    \begin{center}
        \fbox{\includegraphics[width=0.9\textwidth]{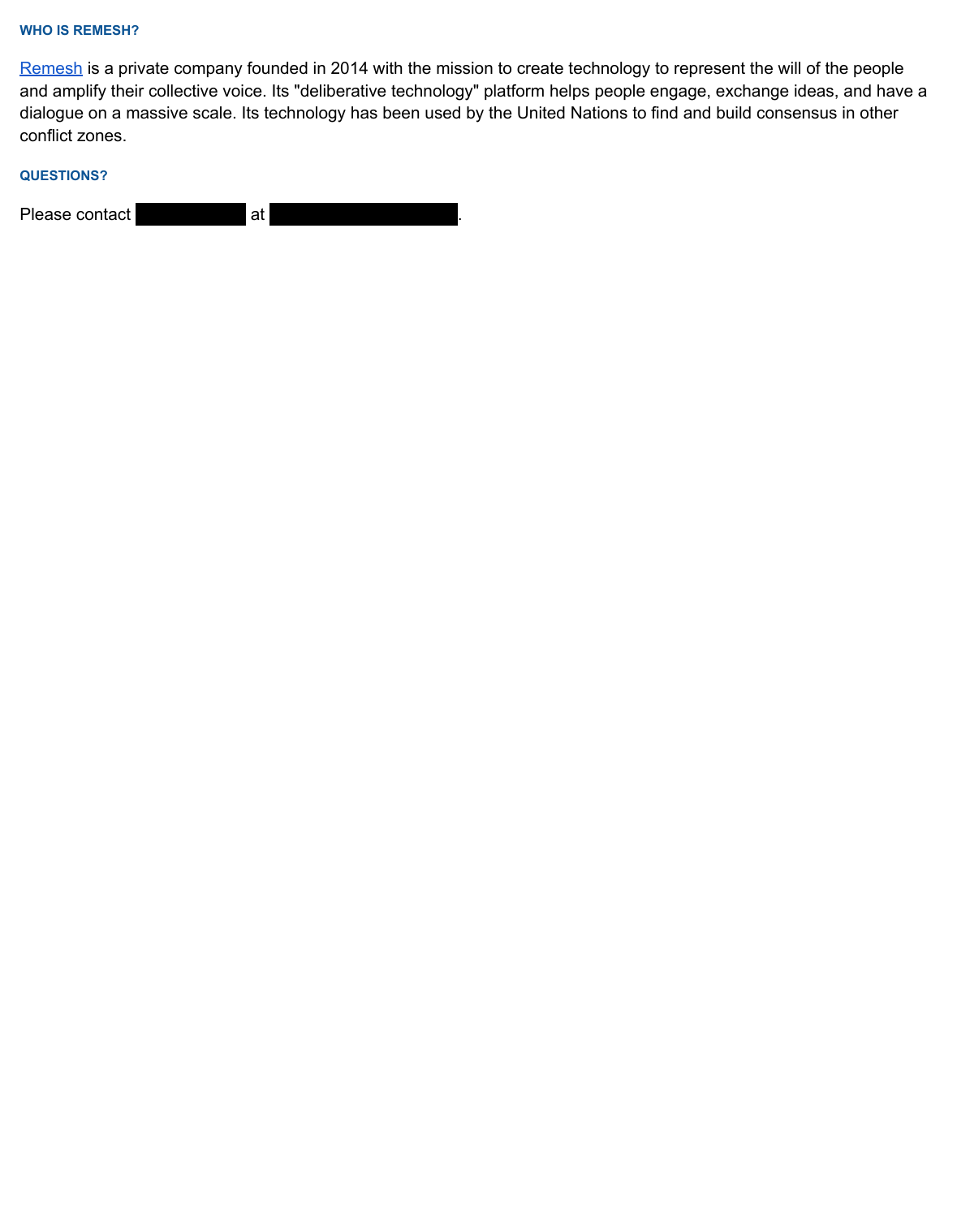}}
    \end{center}
    \clearpage

\section{Sample Sizes}
\label{appendix:counts}

    \begin{table}[H]
        \caption{The tables below record what we know about the number of participants in each dialogue cycle. Columns ``CDa'', ``CDb'', ``CDc'' respectively contain (a) the number who joined the collective dialogue and did demographic onboarding, (b) the number who simultaneously contributed a statement to the same open-ended prompt, and (c) the number contributed statements and voted all the way through to the last collective prompt. Columns ``Va'', ``Vb'', ``Vc'' respectively contain (a) the number who joined and did demographic onboarding for the asyncronous vote, (b) the number who additionally voted on all collective statements in the paper (joint) or on 50\% of collective statements (uninational), and (c) voted on all collective statements not shared in the paper related to internal peacebuilding strategy. Finally, column ``MUP'' contains the maximum possible number unique participants of any sort during the dialogue cycle (Va + CDa), and ``UEP'' contains the best estimate of unique engaged participants during the dialogue cycle (max(CDb, Vb)).}
        \label{table:sample-sizes}

        \includegraphics[width=.75\textwidth]{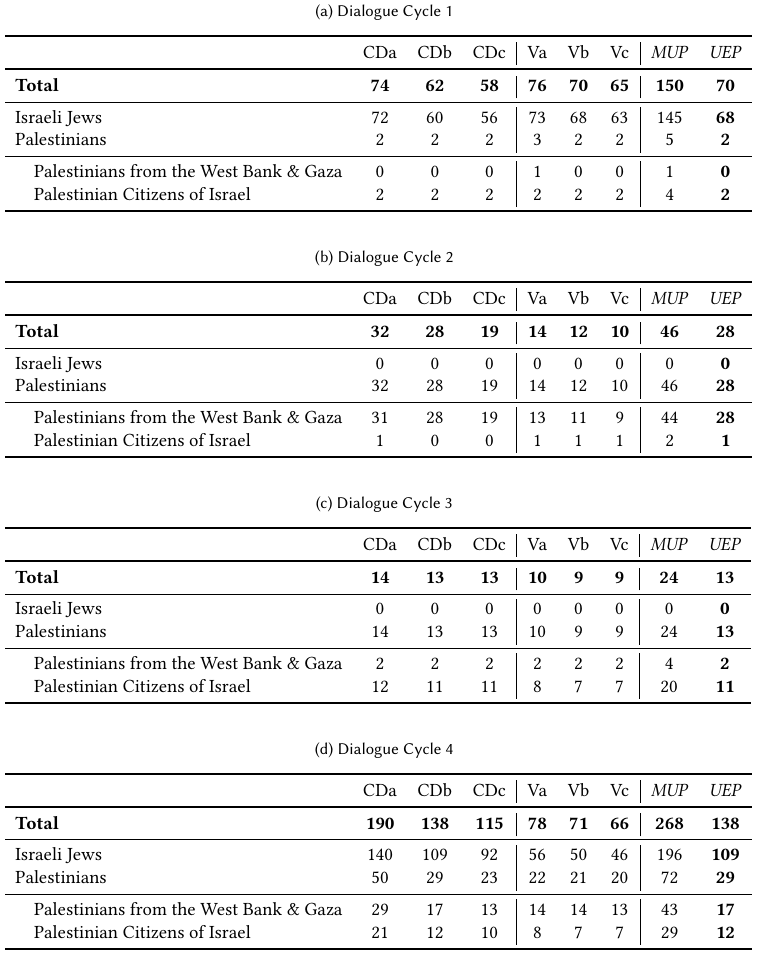}
    \end{table}
    \clearpage

\section{Prompts}
\label{appendix:prompts}

\subsection{Used for Dialogue Cycle 1 in Hebrew}

    \noindent
    \includegraphics[width=\textwidth]{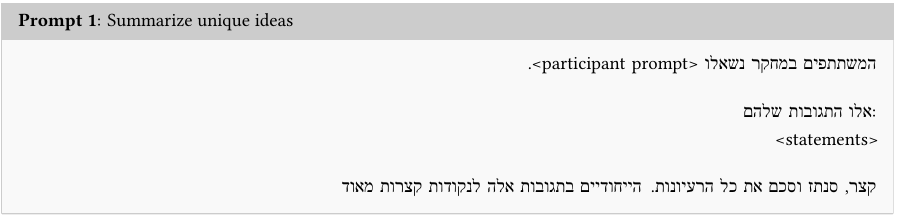}

    \noindent
    \includegraphics[width=\textwidth]{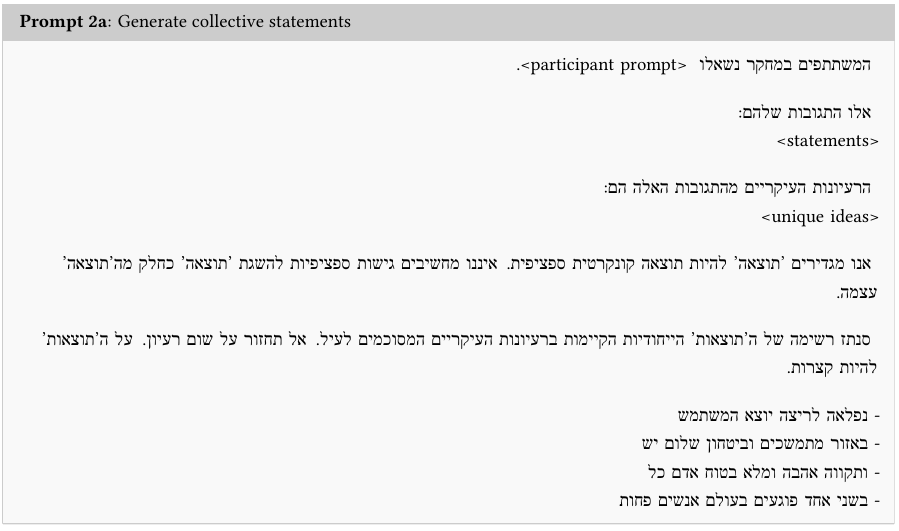}

    \noindent
    \includegraphics[width=\textwidth]{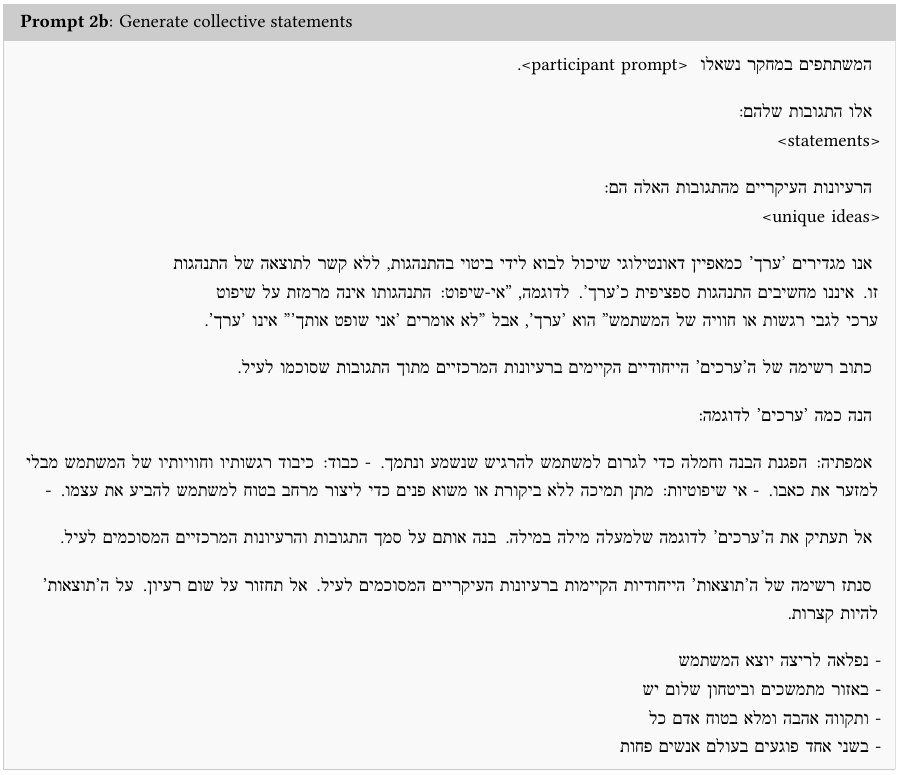}
    
\subsection{Used for Dialogue Cycles 2 and 3 in Palestinian Arabic}

    \noindent
    \includegraphics[width=\textwidth]{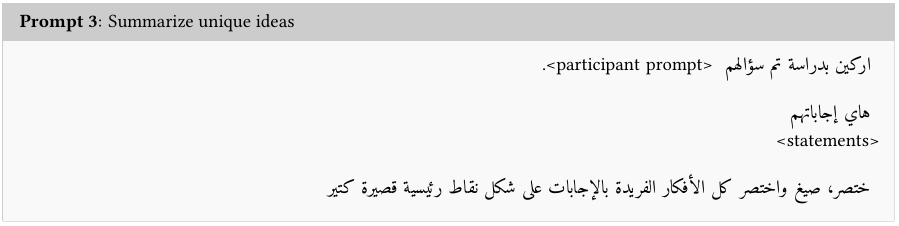}

    \noindent
    \includegraphics[width=\textwidth]{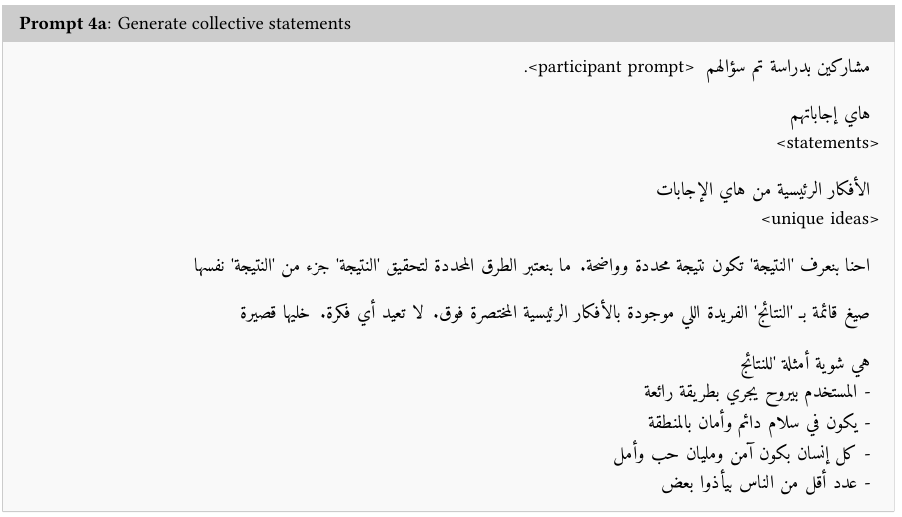}

    \noindent
    \includegraphics[width=\textwidth]{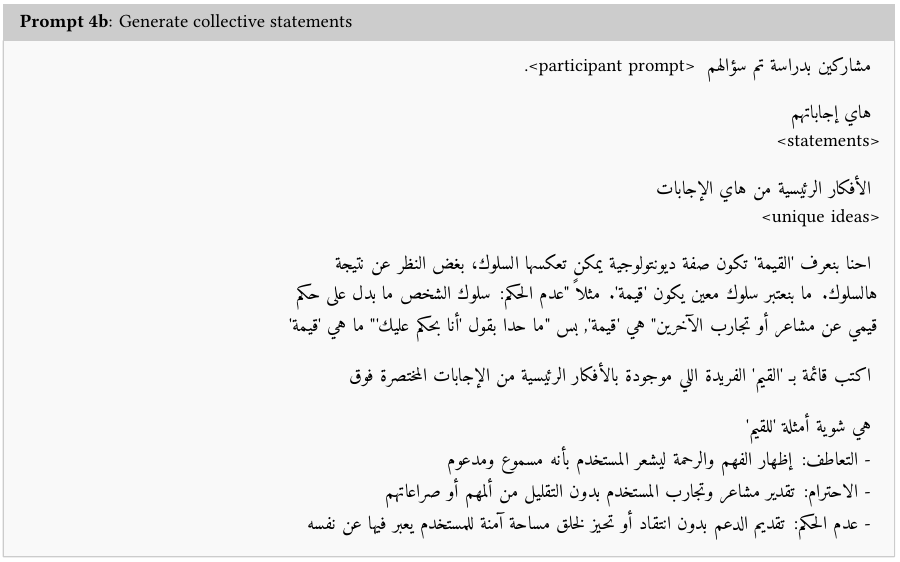}
    
\subsection{Used for Dialogue Cycle 4 in Palestinian Arabic, Hebrew, and English}

    \noindent
    \includegraphics[width=\textwidth]{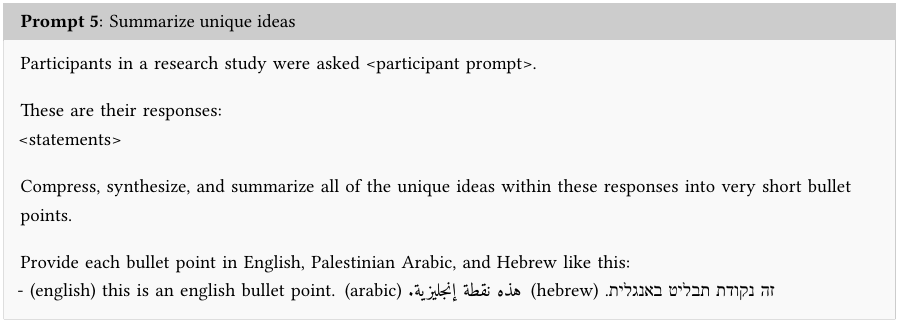}

    \noindent
    \includegraphics[width=\textwidth]{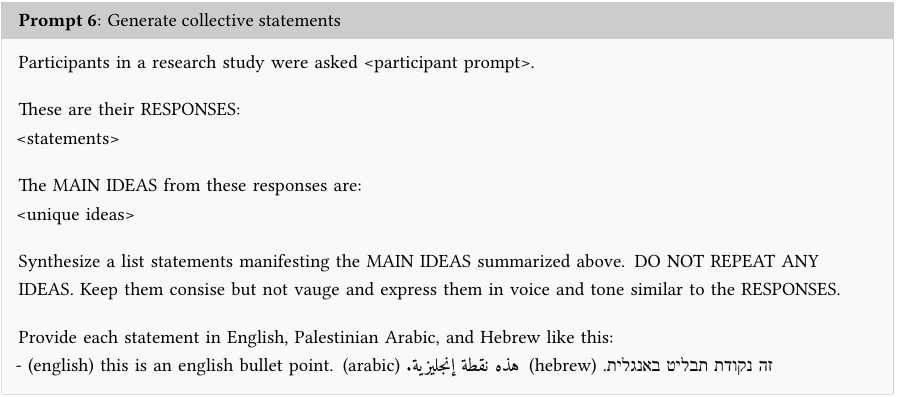}

\end{document}